\documentclass[fleqn,usenatbib]{mnras}
\usepackage{newtxtext,newtxmath}
\usepackage[T1]{fontenc}
\usepackage{graphicx}	% Including figure files
\usepackage{amsmath}	% Advanced maths commands
\newcommand{\alf}{Alfv$\acute{\text{e}}$n } 
\newcommand\bb[1]{\mbox{\boldmath{$#1$}}}

\newcommand{\D}[2]{\frac{{\rm d} #2}{{\rm d} #1}}
\newcommand{\msb}[1]{\bb{\mathsf{#1}}}
\newcommand\grad{\bb{\nabla}}
\newcommand\bcdot{\,\bb{\cdot}\,}
\newcommand\btimes{\,\bb{\times}\,}
\newcommand\eb{\hat{\bb{b}}}
\newcommand\ez{\hat{\bb{z}}}
\newcommand\rmd{{\rm d}}

\begin{document}

\title[Two-temperature accretion disc coronae]{Local models of two-temperature accretion disc coronae. I.~Structure, outflows, and energetics}
\author[C.~J.~Bambic, E.~Quataert and M.~W.~Kunz]  
    {\parbox[]{7.in}{Christopher~J.~Bambic$^{1}$\thanks{E-mail: 
          cbambic@princeton.edu}, Eliot~Quataert$^1$ and Matthew~W.~Kunz$^{1,2}$ \\
    \footnotesize 
    $^1$ Department of Astrophysical Sciences, Peyton Hall, Princeton University, Princeton, NJ 08544, USA \\
    $^2$ Princeton Plasma Physics Laboratory, PO Box 451, Princeton, NJ 08543, USA \\ 
  }
}

\maketitle

\begin{abstract}
We use local stratified shearing-box simulations to elucidate the impact of two-temperature thermodynamics on the thermal structure of coronae in radiatively efficient accretion flows. Rather than treating the coronal plasma as an isothermal fluid, we use a simple, parameterized cooling function that models the collisional transfer of energy from the ions to the rapidly cooling leptons. Two-temperature models naturally form temperature inversions, with a hot, magnetically dominated corona surrounding a cold disc. Simulations with net vertical flux (NF) magnetic fields launch powerful magnetocentrifugal winds that would enhance accretion in a global system. The outflow rates are much better converged with increasing box height than analogous isothermal simulations, suggesting that the winds into two-temperature coronae may be sufficiently strong to evaporate a thin disc and form a radiatively inefficient accretion flow under some conditions. We find evidence for multiphase structure in the corona, with broad density and temperature distributions, and we propose criteria for the formation of a multiphase corona. The fraction of cooling in the surface layers of the disc is substantially larger for NF fields compared to zero net-flux configurations, with moderate NF simulations radiating ${\gtrsim}30$~per cent of the flow's total luminosity above two midplane scale-heights. Our work shows that NF fields may efficiently power the coronae of luminous Seyfert galaxies and quasars, providing compelling motivation for future studies of the heating mechanisms available to NF fields and the interplay of radiation with two-temperature thermodynamics. 
\end{abstract}


\section{Introduction}
\label{sec:intro}

Hard X-ray emission from active galactic nuclei (AGN) accreting at a few percent of the Eddington rate is thought to originate in a hot (electron temperature $T_e \approx 0.2-8 \times 10^9$~K), radiatively compact \citep{Fabian2015}, magnetically dominated structure referred to as the `corona' \citep{Liang1977, Bisnovatyi1977}. The spectral signatures of coronae are remarkably simple: a power-law continuum with photon index $\Gamma \approx 2$ and a spectral cutoff at energy $E_{\rm cut} \approx 20-700$~keV \citep{Turner1989, Mushotzky1993, Haardt1993}. Yet, despite this simplicity, reconciling the unique spectral shape and the high fraction of the total luminosity radiated by the corona with features owing to a coexisting accretion disc has made understanding the formation and maintenance of AGN coronae a formidable theoretical challenge.

The most widely accepted model for the hard emission is thermal Comptonization \citep{Shapiro1976, Sunyaev1980, Zdziarski1999}. Seed photons in the ultraviolet (UV) and soft X-ray bands produced by a ${\sim}10^{4-5}$~K accretion disc are repeatedly Compton upscattered to form a power-law radiation spectrum \citep{Kompaneets1957, Rybicki_Lightman1986}. Electron (or, since pairs may be present, lepton) temperatures $\gtrsim$100~keV provide cutoff energies in accord with observations, and the hard photon indices can be explained by optical depths to electron scattering $\tau_{\rm es} \sim 1$. 

While the geometry of the corona remains highly uncertain \citep{Wilkins2012}, observations are often interpreted within the framework of the two-phase model of \cite{Haardt1991}. In this picture, a hot, diffuse, optically thin corona `sandwiches' a cold, geometrically thin, optically thick accretion disc, providing the optical and UV emission expected of a thin accretion disc \citep{Koratkar1999} alongside the hard X-ray continuum associated with the corona. Observations of radio-quiet Seyfert galaxies indicate the presence of fluorescence from a relativistically broadened iron K$\alpha$ line \citep{Tanaka1995,Fabian2000_FeLines,Reynolds2003} at the innermost stable circular orbit (ISCO) of the black hole \citep{Reis2010}. A two-phase model that presupposes a thin accretion disc extending to the ISCO accounts for this feature. 

Within the context of this observationally motivated model, this paper explores the interplay between a cold, geometrically thin accretion disc and a hot, diffuse corona. Specifically, we focus on the thermodynamics of the corona: the effect of thermodynamics on density, temperature, and magnetic-field structure in the corona; the launching of coronal winds; and energy transport into and out of the corona. While considerable attention has been granted to the dynamics of the disc--corona interaction, understanding the thermodynamics of this system requires investigating the species-dependent mechanisms for cooling coronal plasma: Compton cooling in the case of electrons and positrons, and, in the case of ions, slow transfer of energy to leptons via Coulomb collisions. This work focuses on the latter, namely, how weak ion cooling in the surface layers of AGN discs leads to the formation of a two-temperature corona.

Two-temperature effects may have a number of consequences for global disc evolution and the formation of coronae. In some models of corona formation, thermally unstable, viscous heating evaporates the tenuous surface layers of AGN discs while heating plasma to the virial temperature \citep{Spruit2002}. This evaporation can only occur if ions are decoupled from rapidly Compton-cooled leptons. If ion--lepton collisions are infrequent, then hot, virialized ions must themselves be weakly collisional or collisionless.  High thermal conductivity along magnetic-field lines can channel energy from the corona to the disc, further accelerating the evaporation/coronal-formation process. Two-temperature cooling and conduction may mediate state transitions, at least in X-ray binaries where they are known to occur \citep{Esin1997}, and regulate density in the corona, determining the optical depth as well as the rate of magnetic reconnection in the optically thin plasma \citep{Goodman2008}.

In this paper, we explore a simple model for AGN coronal ions: a single-fluid, magnetohydrodynamic (MHD) plasma with an ideal equation of state (EOS), in which the ions are cooled by an optically thin cooling function that describes the collisional energy transfer from the ions to the rapidly cooling leptons. We examine the implications of our model for disc structure, outflows, and energetics via a suite of carefully controlled, local disc simulations using the stratified shearing-box formalism \citep{Stone1996}. 

The paper is organized as follows. We provide observational motivation for our model in \S\ref{sec:thermo_timescales}. The model itself is described in \S\ref{sec:two_temp_model}. In \S\ref{sec:methods}, we detail a suite of MHD simulations performed in the stratified shearing box, which enable us to study how our cooling function affects the structure, outflows, and energetics of local models of accretion disc coronae. Sections \ref{sec:flow_structure}--\ref{sec:Poynting_cooling} provide our main results. In \S\ref{sec:flow_structure}, we study the thermal structure of our simulated discs: the emergence of temperature inversions, thermally driven winds, the formation of a `multiphase' corona, and estimates of the efficiency of field-aligned conduction between the disc and corona. Magnetocentrifugal winds, which are launched by net-flux magnetic fields and loaded by thermal effects, are examined in \S\ref{sec:wind_accretion}. In \S\ref{sec:Poynting_cooling}, we study the energy transport mechanisms operating in our model: Poynting fluxes into the corona and wind, and cooling rates in the corona. We discuss the observational implications of our findings in \S\ref{sec:discussion} and conclude in \S\ref{sec:conclusion}.

\section{Thermodynamic Timescales}
\label{sec:thermo_timescales}

Observations enabled by missions like the Burst Alert Telescope (BAT) on-board the Neil Gehrels \textit{SWIFT} Observatory \citep{Gehrels2004}, the INTErnational Gamma-Ray Astrophysics Laboratory \citep[\textit{\mbox{INTEGRAL}};][]{Winkler2003}, and the Nuclear Spectroscopic Telescope ARray \citep[\textit{NuSTAR};][]{Harrison2013} indicate that AGN coronae are likely composed of two-temperature plasma. The timescale for ions and leptons to equilibrate their temperatures is much longer than the Compton cooling timescale of the leptons. Here, we provide a comparison of thermodynamic timescales based on the definitions presented in \cite{Fabian2015}, whose observations inspired this work.

Properties of coronae are traditionally constrained by measuring the radiative compactness parameter $\ell$ and the electron (lepton) temperature $\Theta_e \equiv k_{\rm B} T_e/m_e c^2$, where $m_e c^2$ is the electron rest mass. Compactness is defined by $\ell \equiv (\sigma_{\rm T}/m_e c^3) L_{\rm X}/R = 4 \pi (m_i/m_e) (R_{\rm g}/R)(L_{\rm X}/L_{\rm Edd})$ \citep{Guilbert1983}, where $L_{\rm X}$ is the $0.1$--$200$~keV X-ray flux (i.e., that originating in the `corona'), $L_{\rm Edd} \equiv 4 \pi {\rm G} M_{\rm BH} m_i c/\sigma_{\rm T}$ is the Eddington luminosity, and $R$ is the `size' of the corona. Coronae are thought to form within $3$--$10$ gravitational radii $R_{\rm g} \equiv G M_{\rm BH}/c^2$ of a supermassive black hole (SMBH) with mass $M_{\rm BH}$, as motivated by X-ray timing observations \citep{Fabian2009, deMarco2011, Kara2013, Cackett2014, Emmanoulopoulos2014, Uttley2014}.

Hard X-ray observations indicate that AGN coronae have large radiative compactnesses \citep[$\ell \gtrsim 10$; ][]{Fabian2015, Kamraj2022} and a fairly limited range of electron (lepton) temperatures imposed by efficient Compton cooling ($\Theta_e \approx 0.03 - 1.36$). This limited range may be in part a result of photon--photon pair production, which may regulate $\Theta_e$ through a `pair thermostat' \citep{Svensson1984, Zdziarski1985, Pietrini1995, Stern1995, Coppi1999, Dove1997}, with excess lepton heating going to the production of pairs rather than an increase in $\Theta_e$.

Large radiative compactness implies that the Compton cooling time $t_{\rm IC}$ in the corona is extraordinarily short,
\begin{equation} \label{eq:IC_time}
    t_{\rm IC} \lesssim 2 \times 10^2 \: \frac{1}{1+\tau_{\rm es}} \left( \frac{\ell}{10} \right)^{-1} \left( \frac{R}{10 \: R_g} \right) \left( \frac{M_{\rm BH}}{10^7 \: \rm{M}_{\odot}} \right) \: \rm{s}.
\end{equation}
Here, we assume that the radiation energy density in the corona is $\mathcal{E}_{\rm rad} \equiv (1 + \tau_{\rm es}) (L_{\rm X}/4 \pi R^2 c)$, where $R = 10 R_{\rm g}$ (an estimate we use throughout this work). The Keplerian orbital timescale is $t_{\rm orb} \equiv 2\pi/\Omega \simeq 9.8 \times 10^3 ( M_{\rm BH}/10^7 \: {\rm M}_{\odot} ) \: \rm{s}$ at $10 R_{\rm g}$, where $\Omega$ is the orbital frequency and we have ignored general relativistic effects. Thus, any energy gained by leptons is immediately radiated away on timescales far shorter than the disc dynamical time or even the light-crossing time of the hole.

Unlike electrons, ions are subject to much more gradual cooling. Coulomb collisions slowly transfer energy from ions to rapidly cooling electrons on the ion--electron thermal equilibration timescale,
\begin{equation} \label{eq:temp_equilibration}
\begin{split}
    t_{\rm eq} &= \frac{1}{2} \frac{m_i}{m_e} \frac{3 \sqrt{m_e} (k_{\rm B} T_e)^{3/2}}{4 \sqrt{2\pi} n_e e^4 \ln{\Lambda_e}} \\
    &\approx 
    5 \times 10^3 \: \left(\frac{\Theta_e}{0.2} \right)^{3/2} 
    \left( \frac{n_e}{10^{11} \: \rm{cm}^{-3}} \right)^{-1} \left( \frac{\ln{\Lambda_e}}{23} \right)^{-1} \: \rm{s}.
\end{split}
\end{equation}
Here, we have introduced the electron Coulomb logarithm $\ln{\Lambda_e}$ and the electron number density $n_e$. In thermal Comptonization models, the observed photon index and cutoff energy equate to a Thomson optical depth and electron temperature, respectively, with $\tau_{\rm es}$ usually ${\sim}\mathcal{O}(1)$. Using the simple approximation for the path length integral over the corona, $\tau_{\rm es}~\sim~n_l \sigma_T R~\sim~1$, we can obtain a rough estimate for the electron number density, $n_e \approx 10^{11} ( M_{\rm BH}/ 10^{7} M_{\odot} )^{-1} \: \rm{cm}^{-3}$. Equation \ref{eq:temp_equilibration}
%then
shows that the ion--electron equilibration timescale 
far exceeds %the local dynamical timescale and 
the Compton cooling timescale. The coronal plasma is thus \textit{two-temperature}, with the ion temperature $T_i$ exceeding the electron temperature $T_e$ \citep{DiMatteo1997, Goodman2008}. 

The ion--ion collision time $t_{ii}$ is longer than the equilibration time by a factor of ${\sim}(T_i/T_e)^{3/2} (m_e/m_i)^{1/2}$, where we have ignored the (weak) contribution of the Coulomb logarithm. If ions are virialized, with temperature $k_{\rm B} T_{\rm virial} \equiv (1/3) m_i c^2 (R/R_{\rm g})^{-1}$, then at $10 R_{\rm g}$, the ion temperature $k_{\rm B}T_i~\approx 31~\rm{MeV}$. For 100 keV electrons, ion--ion collisions occur on a timescale ${\approx}130$ times longer than $t_{\rm eq}$. Thus, AGN coronae should obey the ordering $t_{ii} \gg t_{\rm eq} \simeq t_{\rm orb} > t_{\rm IC} \gtrsim t_{ee}$, where $t_{ee}$ is the timescale over which electrons exchange momentum. 

\section{Two-Temperature model}
\label{sec:two_temp_model}

In this paper, we present a single-fluid, MHD model for the plasma within an ion-supported corona. Rather than study ions through an isothermal or nearly isothermal EOS, as is the case in most previous studies \citep{Stone1996, Miller2000, Bai2013_NF, Zhu2018, Mishra2020}, we assume that the ions obey an ideal equation of state subject to an explicit cooling term that models the Coulomb-collisional transfer of energy from the ions to the colder electrons. Because the ion pressure greatly exceeds the electron pressure in a two-temperature corona, it is reasonable to first explore a single-fluid model focusing on ion dynamics.

Ions in the corona are cooled via Coulomb collisions with electrons. Energy leaves the ions at a rate set by the equilibration time,
\begin{equation}\label{eq:cooling}
    \D{t}{\mathcal{E}_{\rm int}} = - \frac{\mathcal{E}_{\rm int}}{t_{\rm eq}}.
\end{equation}
Here, the internal energy $\mathcal{E}_{\rm int} = (3/2) n_i k_{\rm B} T$, where $n_i$ is the ion number density and $T$ is the MHD fluid (ion) temperature, with $T~\gg~T_e$ \citep{Goodman2008}. Since the Coulomb logarithm has only a weak dependence on density and temperature, the equilibration time obeys the approximate scaling, $t_{\rm eq} \propto n_e^{-1} \Theta_e^{3/2}$. Thus, the rate of cooling experienced by the ions should scale as
\begin{equation} \label{eq:scaling}
    \D{t}{\mathcal{E}_{\rm int}} \propto - \frac{n_i n_e T}{\Theta_e^{3/2}}.
\end{equation}
Motivated by efficient regulation of $T_e$ by Compton cooling (and potentially, pair production), we assume that $\Theta_e$ does not vary significantly over the corona; although, a full radiative-transfer calculation will ultimately be necessary to verify or refute this assumption.

For use in our simulations, we express the cooling function \eqref{eq:cooling} in terms of the fluid density $\rho$ and temperature $T$, which are scaled by their values at the disc midplane, $\rho_0$ and $T_0$. By assuming quasi-neutrality ($n_e = n_i$), the Coulomb cooling function takes the form,
\begin{equation} \label{eq:cooling_function}
    \D{t}{\mathcal{E}_{\rm int}} = -\Lambda (\rho, T) = -2 {\mathcal{A}} \left( \frac{\rho}{\rho_0} \right)^2 \left( \frac{T}{T_0} \right) \: \rho_0 T_0 \Omega \:\:{\rm for}\:\: T > T_0,
\end{equation}
where $k_{\rm B} = m_i = 1$ and the uncertain electron physics is absorbed into a constant free parameter $\mathcal{A}$ that we vary in our model. Throughout this work, we refer to $\mathcal{A}$ as the Coulomb coupling parameter. Note that, for reasons explained below, the two-temperature cooling function \eqref{eq:cooling_function} has a floor imposed at $T = T_0$. 

Equating the physical ratio $\mathcal{E}_{\rm int}/t_{\rm eq}$ with our two-temperature cooling function $\Lambda$ (Equation~\ref{eq:cooling_function}) provides an expression for $\mathcal{A}$,
\begin{equation} \label{eq:A_Omega_teq}
    \mathcal{A} = \frac{3}{4} \left( \frac{n_{i,\rm c}}{n_0} \right)^{-1} \frac{1}{\Omega t_{\rm eq}}.
\end{equation}
Here, $n_{i, \rm c}$ is the typical number density of \textit{ions} in the corona, and $n_0$ is the number density in the disc. Crucially, although observations do not constrain $n_{i, \rm c}/n_0$, they do constrain $\Omega t_{\rm eq}$: 
\begin{equation} \label{eq:Omega_teq_obs}
    \Omega t_{\rm eq} \simeq 7 \: \chi \: \tau_{\rm es}^{-1} \left( \frac{\Theta_e}{0.2} \right)^{3/2} \left( \frac{\ln{ \Lambda_e }}{23} \right)^{-1} \left( \frac{R}{10 \: R_{\rm g}} \right)^{-1/2},
\end{equation}
where $\chi$ is an $\mathcal{O}(1)$ (possibly $\mathcal{O}(H/R)$) factor set by the geometry of the corona. Note that Equation~\ref{eq:Omega_teq_obs} holds even if the corona is pair-dominated, i.e., the lepton density in the corona $n_{l, \rm c} \gg n_{i, \rm c}$.

We study three different values of the Coulomb coupling parameter, $\mathcal{A} \in \{ 10, 100, 10^3 \}$. These values are chosen for the following reasons. The coronal density in our simulations is generally ${\ll}n_0$, the midplane density. Values of $\mathcal{A} \sim 10$--$10^3$ are sufficiently large to regulate the midplane temperature at the minimum of our cooling function (Equation~\ref{eq:cooling_function}), ${\sim}T_0$, i.e., to keep it roughly isothermal. At the same time, these $\mathcal{A}$ values are sufficiently small  that, in the surface layers above a few scale heights, the cooling time ($t_{\rm eq}$ for the ions) satisfies the inequality $t_{\rm eq} \Omega \gtrsim 1$, consistent with observational inferences about coronal properties summarized in \S\ref{sec:thermo_timescales}. Further motivation for our choices of $\mathcal{A}$ can be found by studying global simulations.

Recent global simulations of radiating accretion flows---performed either by solving the radiation-transfer equation directly while neglecting general relativistic (GR) effects \citep{Jiang2019}, or by including GR but computing the radiation transport using the M1 closure approximation \citep{Fragile2012, Fragile2014, McKinney2014, Sadowski2016, Liska2019, Liska2020, Liska2022}---have begun to provide insight into the thermodynamic structure of radiatively efficient accretion flows. To get a sense for which values of $\mathcal{A}$ are physically relevant, we consider the simulation presented by \cite{Jiang2019} of an AGN with black-hole mass $5\times 10^8~{\rm M}_\odot$ accreting at $7$ per cent of the Eddington rate. Those authors found that the ratio of the coronal density $n_{i, \rm c}$ to the midplane density $n_0$ is $n_{i, \rm c}/n_0 \approx 10^{-2}$--$2\times 10^{-4}$ (see their figure~10). Adopting an intermediate value of $n_{i, \rm c}/ n_0 = 2\times 10^{-3}$, we find that our chosen values of $\mathcal{A} \in \{ 10, 100, 10^3 \}$ correspond to equilibration times $\Omega t_{\rm eq} \in \{ 40, 4, 0.4 \}$. Thus, our simulations probe the regime from strong Coulomb coupling, with $\Omega t_{\rm eq} \lesssim 1$, to weak Coulomb coupling, with $\Omega t_{\rm eq} \gtrsim 1$. 

\section{Methods}
\label{sec:methods}

\subsection{Equations solved}
\label{eq:equations_solved}

We use the \textit{Athena++} MHD code \citep{Stone2020} to evolve the equations of ideal MHD in a coordinate system centred at a radial distance $R_0$ from the black hole and corotating with a local patch of plasma at the Keplerian angular velocity $\boldsymbol{\Omega} = \Omega \ez$. Because the shearing box is a local approximation, global coordinates $(r,\varphi,z)$ are mapped to Cartesian coordinates $(x,y,z)$ through $x = r - R_0$, $y = R_0 \varphi - \Omega t$, and $z = z$, where the simulation evolves with time $t$. This approximation applies when the radial extent of the domain $L_x \ll R_0$ such that curvature terms can be ignored.

In the frame corotating with the disc, the ideal MHD equations including cooling and a source (`src') term to compensate for mass lost by the shearing box ($\dot{\rho}_{\rm src}$) are given in conservative form as
\begin{gather} \label{eq:mass}
    \frac{\partial \rho}{\partial t} + \grad \bcdot \left( \rho \boldsymbol{u} \right) = \dot{\rho}_{\rm src},
\end{gather}
\begin{gather} \label{eq:momentum}
\begin{split}
    &\frac{\partial}{\partial t} \rho \boldsymbol{u} + \grad \bcdot \left[ \rho \boldsymbol{u} \boldsymbol{u} - \boldsymbol{B} \boldsymbol{B} + \left(P + \frac{|\boldsymbol{B}|^2}{2}\right) \msb{I} \right] \\ 
    \mbox{} &\quad =  \rho \left( 2 \boldsymbol{u} \btimes \boldsymbol{\Omega} + 3 \Omega^2 \boldsymbol{x} \right)
    - \rho \grad \Phi + \dot{\rho}_{\rm src} \boldsymbol{u},
\end{split}
\end{gather}
\begin{gather} \label{eq:energy}
\begin{split}
    &\frac{\partial \mathcal{E}}{\partial t} + \grad \bcdot \left[ \left( \mathcal{E} + P + \frac{|\boldsymbol{B}|^2}{2} \right) \boldsymbol{u} - \boldsymbol{B} \left( \boldsymbol{B} \bcdot \boldsymbol{u} \right) \right]
    \\
    \mbox{} &\quad = 3\rho \Omega^2 \boldsymbol{u} \bcdot \boldsymbol{x}   
    - \rho \boldsymbol{u} \bcdot \grad \Phi
    + \dot{\rho}_{\rm src} \left( \frac{T}{\gamma - 1} + \frac{|\boldsymbol{u}|^2}{2} \right)
    - \Lambda (\rho,T),
\end{split}
\end{gather}
\begin{gather} \label{eq:induction}
    \frac{\partial \boldsymbol{B}}{\partial t} = \grad \btimes \left( \boldsymbol{u} \btimes \boldsymbol{B} \right).
\end{gather}
Here, $\rho, \boldsymbol{u}$, and $P$ are the fluid density, velocity, and pressure respectively; $\boldsymbol{B}$ is the magnetic field with factors of $\sqrt{4 \pi}$ absorbed; $\msb{I}$ is the rank-2 identity tensor; and $\Phi$ is the gravitational potential. By expanding the potential about a point $(r, z) = (R_0, 0)$ in the disc, we find that the vertical contribution to the local gravitational potential is
\begin{equation} \label{eq:gravitational_potential}
    \Phi = \frac{1}{2} \Omega^2 z^2.
\end{equation}
The energy density $\mathcal{E}$ is given by
\begin{equation} \label{eq:energy_density}
    \mathcal{E} = \frac{1}{2} \rho |\boldsymbol{u}|^2 + \frac{|\boldsymbol{B}|^2}{2} + \frac{P}{\gamma - 1},
\end{equation}
where $\gamma$ is the adiabatic index such that the plasma obeys an ideal EOS, $P = (\gamma - 1) \mathcal{E}_{\rm int}$, with $\gamma = 5/3$. Equating $\mathcal{E}_{\rm int} = \rho \Phi$ yields a virial temperature for a simulation domain with maximum vertical extent $z_{\rm max} = L_z/2$, appropriate for our shearing-box simulations,
\begin{equation} \label{eq:T_virial}
    T_{\rm virial} = \frac{\gamma - 1}{2} \Omega^2 z_{\rm max}^2 = \frac{1}{3} \Omega^2 z_{\rm max}^2.
\end{equation}
In \S\ref{sec:hot_corona}, we discuss how this definition allows us to connect the temperatures in our simulations to the more standard spherical virial temperature introduced in \S\ref{sec:thermo_timescales}.

For this initial investigation, we consider an inviscid, perfectly conducting fluid such that non-ideal terms in the MHD equations vanish. This assumption removes transport terms that arise from the weakly collisional nature of the corona, including viscosity and thermal conduction. In the absence of cooling and the source term, Equations~\eqref{eq:mass}--\eqref{eq:induction} are conservative. Any energy dissipated in this system, which occurs purely through grid diffusion since we do not implement explicit viscosity or resistivity, is not lost by the system. 

The cooling function~\eqref{eq:cooling_function} is implemented as a source term in the energy equation~\eqref{eq:energy} and calculated in an operator-split fashion using the exact integration scheme devised by \cite{Townsend2009}. 

\subsection{Initial conditions}
\label{sec:ICs}

We work in units in which the local Keplerian orbital frequency is $\Omega=1$, the midplane density $\rho_0 = 1$, and the midplane temperature $T_0 = 1/2$. The latter implies an isothermal speed of sound at the midplane $c^2_{\rm s0} = T_0$ that is related to the thermal scale height of the disc $H_z$ through vertical force balance:
\begin{equation} \label{eq:Hz}
    H_z^2 = \frac{2 c_{\rm s0}^2}{\Omega^2} = 1.
\end{equation}
Hydrostatic equilibrium with the local gravitational potential corresponds to a density distribution,
\begin{equation} \label{eq:density_profile}
    \rho(z) = {\rm max} \left( \rho_0 {\rm e}^{-z^2/H_z^2}, \: \rho_{\rm floor} \right).
\end{equation}
In arriving at this expression, we assumed the plasma is initially at a single temperature $T_0$. The pressure profile corresponding to this density distribution is
\begin{equation} \label{eq:pressure_profile}
    P(z) = {\rm max} \left( P_0 {\rm e}^{-z^2/H_z^2}, \: P_{\rm floor} \right),
\end{equation}
where $P_0 = \rho_0 c_{\rm s0}^2 = 0.5 \rho_0 H^2_z\Omega^2$ is the midplane pressure, and $\rho_{\rm floor}$ and $P_{\rm floor}$ are the floor density and pressure (\S\ref{sec:floors}), respectively. Our simulations are initialized with these profiles. Throughout this work, time is measured relative to the orbital timescale $t_{\rm orb} = 2 \pi/ \Omega = 2\pi$.

The initial magnetic-field configuration is described in terms of the net vertical magnetic flux threading the disc, $\Phi_{\rm B} = \oiint \boldsymbol{B} \bcdot \hat{\bb{z}} \: \rmd x \rmd y$. Net vertical flux is conserved in the stratified shearing box: it can neither be created nor destroyed. For a field configuration with zero net flux (ZNF), we implement the field of \cite{Miller2000}, specified in terms of the vector potential $\boldsymbol{A} = A_x \hat{x}$ with
\begin{align} \label{eq:vector_potential}
    A_x(x,y,z) &= \sqrt{\frac{2 P_0}{\beta_0}} \cos{\left( 2 \pi x \right)} 
    \cos{ \left( y \right)} \nonumber\\*
    \mbox{} &\times\, 
    \begin{cases}
    ~1, & |z| \leq 2 H_z \\
    ~\exp\bigl[(|z| - 2 H_z)^4\bigr], & 2H_z < |z| < 4 H_z \\
    ~0, & |z| \geq 4 H_z.
    \end{cases}
\end{align}
The magnetic field $\bb{B} = \grad \btimes \bb{A}$ is initialized by finite differencing this vector potential to ensure that $\boldsymbol{\nabla} \bcdot \boldsymbol{B} = 0$. 

Our motivation for this choice of ZNF field is two-fold: (1) \cite{Miller2000} found that this configuration yields a substantial Poynting flux into the corona, enough to heat the rarefied extra-planar regions to temperatures consistent with X-ray observations; and (2) these field loops are confined to the disc bulk ($|z| \leq 4 H_z$ but super-exponentially attenuated above $|z| = 2 H_z$). Unlike previous works in which magnetic-field loops extending out of the the disc and into the corona are pre-supposed \citep[e.g.,][]{Yuan2019,Chashkina2021,El_Mellah2022}, significant magnetic energy only reaches $|z| > 2 H_z$ in our simulations if MRI-driven turbulence and/or winds self-consistently transport it into the corona.  

Simulations with net vertical flux (NF) are initialized using a vertical field with constant $B_z$. In both the ZNF and NF simulations, the magnetic-field strength is determined by the initial midplane plasma $\beta$ parameter at time $t=0$: $\beta_0 \equiv 2 P_0 (t=0)/|\boldsymbol{B}_0 (t = 0)|^2$. For the ZNF simulations, we study cases where $\beta_0 = 10^2$ (weak ZNF) and $10$ (moderate ZNF), and for NF fields, we analyze cases with $\beta_0 = 10^4$ (weak NF) and $10^3$ (moderate NF). 

% local shearing-box
% Zhu & Stone (2018) in particular
The strengths of the NF fields were chosen to facilitate comparison with previous local simulations by \cite{Fromang2013}, \cite{Lesur2013}, \cite{Bai2013_NF}, and \cite{Salvesen2016_NF}, as well as global calculations by \cite{Zhu2018}, who used a $\beta_0 = 10^3$ vertical field. Using an effectively isothermal treatment of the disc thermodynamics, \cite{Zhu2018} found that stronger fields resulted in rapid accretion and a build-up of vertical flux on the inner radial boundary. In our case, stronger NF fields require $\mathcal{A}$ to be larger than $10^3$ to maintain the disc temperature at ${\sim}T_0$. For ZNF fields, we choose field strengths that are sufficiently subthermal so that the MRI, rather than $\beta_0$, determines the final saturated field strength. In addition, $\beta_0 = 10$ matches the field strength used in the radiation-MHD simulations of \cite{Jiang2014}; the case with $\beta_0 = 10^2$ is used primarily to verify that the saturated magnetic field does not depend on $\beta_0$, so long as $\beta_0$ is sufficiently large.

\subsection{Boundary conditions}
\label{sec:BCs}

Our domain is shearing periodic in the $x$-direction and periodic in the $y$-direction. As is well known with stratified shearing boxes, the structure of winds is affected by the choice of the $z$-boundary condition. \cite{Lesur2013} provides a detailed study of the effects of this boundary condition in local simulations. They find that standard outflow conditions (viz., copying the velocity $\boldsymbol{v}$ and magnetic field $\boldsymbol{B}$ with zero gradient from the last active cell into the ghost zones) suppresses outflows, leading to an unphysical build-up of magnetic energy at large $|z|$ and compression of the disc midplane.  

We therefore adopt modified outflow boundary conditions, as follows. Velocities are copied into the ghost zones subject to a `diode' condition, where outgoing $z$-velocities are copied into ghost zones with zero gradient and in-going velocities are set to 0. The magnetic field obeys the boundary condition of \cite{Lesur2013}: the $y$ and $z$ components of the magnetic field are copied into the ghost zones with zero gradient, but the $x$ component of the field is set to 0 in the ghost zones. This choice maintains horizontal (i.e., in the $x$-$y$ plane) currents at the upper/ lower boundary, allowing the Lorentz force to drive an outflow. 

Pressure and density are exponentially extrapolated into the ghost zones at the $z$-boundaries. These profiles in the ghost zones are set to those in Equations~\ref{eq:density_profile} and \ref{eq:pressure_profile}, modified by replacing $\rho_0$ and $P_0$ with the densities and pressures in the last active zone $\rho(x,y,\pm z_{\rm max})$ and $P(x,y,\pm z_{\rm max})$, and the scale height with $H_z^2 \rightarrow H_z^2 (T/T_0)$, where $T=T(x, y, \pm z_{\rm max})$ is the temperature in the last active zone. We impose the pressure floor (\S\ref{sec:floors}) before computing the temperature.

\subsection{Sources and sinks: turbulence, cooling, and outflows}
\label{sec:sources_sinks}

Shearing-periodic boundary conditions impose a boundary term in the energy equation that injects energy at the rate \citep{Hawley1995}
\begin{equation} \label{eq:turbulent_injection}
    \dot{E}_{\rm turb} = \int Q^{+}_{\rm turb} \: \rmd^3 \boldsymbol{r} = \frac{1}{2} q \Omega L_x \oiint \mathcal{T}_{r \varphi} (x = \pm L_x/2) \: \rmd y \rmd z,
\end{equation}
where $q \equiv -\rmd \ln{\Omega}/\rmd \ln{r} = 3/2$ is the shear parameter evaluated for a Keplerian flow, $L_x$ is the radial extent of the domain (i.e., the length in the $x$-direction), and the surface integral is taken over the shearing boundary. The factor of $1/2$ is because we integrate over \textit{both} shearing boundaries ($x=\pm L_x/2$) and then average the values. 
%Use of the superscript `$+$' implies that this energy is added to the domain. 
The $\mathcal{T}_{r\varphi}$ stress is the $r$-$\varphi$ component of the turbulent stress tensor,
\begin{equation} \label{eq:Trphi}
    \boldsymbol{\mathcal{T}} \equiv \underbrace{\rho \boldsymbol{v} \boldsymbol{v}}_{\rm Reynolds} \: \: \underbrace{- \boldsymbol{B} \boldsymbol{B}}_{\rm Maxwell},
\end{equation}
where the turbulent velocity $\boldsymbol{v}= \boldsymbol{u} + \frac{3}{2} \Omega x \hat{\bb{y}}$ does not include the background Keplerian shear. 
 
 Cooling acts as a sink for injected energy. The cooling rate is  
\begin{equation} \label{eq:cooling_rate}
    \dot{E}_{\rm cool} = \int Q^{-}_{\rm cool} \: \rmd^3 \boldsymbol{r} = \int \Lambda (\rho, T) \: \rmd^3 \boldsymbol{r},
\end{equation}
where the volume integral is taken over the entire domain. 

Because coronae are the launchpads of accretion-disc winds \citep{Stone1996}, our model includes a strong mass loss due to winds. Without compensating for this mass loss, we would never obtain the quasi-steady state solution necessary for our study. Thus, mass, momentum, and energy are all sourced such that
\begin{equation} \label{eq:Mdot_wind}
    \dot{M}_{\rm wind} (t) = \int \dot{\rho}_{\rm src} \: \rmd x \rmd y \rmd z = \oint \rho u_z \: \rmd x\rmd y.
\end{equation}
We inject mass at the rate $\dot{\rho}_{\rm src}$ with the initial Gaussian density profile (Equation~\ref{eq:density_profile}). Mass is injected into a cell with the same velocity $\boldsymbol{u}$ and temperature $T \equiv P/\rho$ as the material already in that cell. Note that this injection differs from the `renormalization' procedure used by \cite{Bai2013_NF} and \cite{Salvesen2016_NF}, who multiply the density in every cell by a common factor to maintain a constant box mass. Our mass injection method is chosen to minimize artificial mass and energy injection into the corona while maintaining a constant box mass, despite outflows. 

\subsection{Domain size and resolution}
\label{sec:domain_resolution}

We use a domain of size $(L_x, L_y, L_z) = (4 H_z, 8 H_z, 12 H_z)$  with a resolution of 32 grid cells per $H_z$ and evolve all simulations for 100 orbits until time $\Omega t \approx 628$. The vertical box height $z_{\rm max} = 6 H_z$ exceeds that of \cite{Jiang2014}, whose formation of a hot corona in a domain with $z_{\rm max} = 4 H_z$ and ZNF fields at 32 cells/$H_z$ resolution with radiation-MHD inspired our work. Additionally, our $z_{\rm max}$ provides useful comparisons to the NF calculations of \cite{Bai2013_NF} and \cite{Salvesen2016_NF}, who used $z_{\rm max} = 6 H_z$ and ${\approx}7H_z$, respectively. We study the effect of box height on mass and energy outflow rates in the NF $\mathcal{A} = 10$, $\beta_0 = 10^4$ simulation in \S\ref{sec:height_convergence}.

The chosen radial and azimuthal extents are motivated by published isothermal, stratified, shearing-box simulations. When the radial extent $L_x$ exceeds 2$H_z$ in ZNF isothermal simulations, the Maxwell and Reynolds stresses more than double, and long-lived (several 10s of orbits) zonal flows develop \citep{Johansen2009}. Similarly, azimuthally extended mesoscale structures are always present at the largest scales of the domain \citep{Guan2011, Simon2012}. These structures play a key role in the formation of buoyant flux ropes and thus energy transport into the corona \citep{Blackman2009}. The large radial ($L_x = 4 H_z$) and azimuthal ($L_y = 8 H_z$) extents used in our work faithfully capture these effects. Convergence of our two-temperature shearing boxes with increasing resolution is at present unclear, and we leave this question, particularly the convergence of energy transport and dissipation in the corona, to future work. 

\subsection{Floors}
\label{sec:floors}

The initial density profile $\rho (t=0) \propto \exp{(-z^2/H^2_z)}$ drops off steeply with height. At $z_{\rm max} = 6 H_z$, $\rho \approx 2 \times 10^{-16} \rho_0$. To enable our simulations to run with reasonable computational expense, we impose a density floor $\rho_{\rm floor} = 10^{-4} \rho_0$ in the NF simulations and $\rho_{\rm floor} = 10^{-6} \rho_0$ in the ZNF simulations. Correspondingly, a pressure floor $P_{\rm floor} = \rho_{\rm floor} T_0$ is imposed in all runs. 

Because NF simulations launch strong outflows, densities in the corona are greatly enhanced relative to ZNF simulations; NF simulations can be performed with a higher density floor relative to ZNF runs. Our chosen density floor in NF runs, $\rho_{\rm floor} = 10^{-4} \rho_0$, is the same as was used in \cite{Salvesen2016_NF}. We find that horizontally averaged densities remain well above the floor in all NF simulations. The exception may be the weak-field isothermal simulation, in which the horizontally averaged density comes within a factor of 2--3 of the density floor, although only near the upper boundary. 

ZNF runs are far more affected by the choice of floor. With the NF floor value of $\rho_{\rm floor} = 10^{-4} \rho_0$, the atmospheres of our $\mathcal{A} = 10^3$ and isothermal ZNF simulations collapsed, evacuating the corona and requiring more floored material to be injected into the simulation. With a floor of $\rho_{\rm floor} = 10^{-5} \rho_0$, we found that instead of collapsing, the $\mathcal{A} = 10^3$ atmospheres developed an outflow. Turning down the floor further maintained this behaviour, with nearly identical vertical velocity profiles. Our chosen floor of $\rho_{\rm floor} = 10^{-6} \rho_0$ is thus low enough to ensure a robust and accurate solution. 

\begin{table*} 
\renewcommand{\arraystretch}{1.1}
\small\addtolength{\tabcolsep}{-2pt}
%\begin{centre}
\scalebox{1}{%
\begin{tabular}{c c c c c c c c c c}     
\hline  
Simulation & $t_{\rm dep}$ & $t_{\rm thm}$  & $\langle \dot{E}_{\rm turb} \rangle_t$ &  $\langle \dot{E}_{\rm wind} \rangle_t$  & $\langle \dot{E}_{\rm cool} \rangle_t$ & $\langle \dot{E}_{\rm turb}^{\rm disc} \rangle_t$ & $\langle \dot{E}_{\rm Poyt}^{\rm cor} \rangle_t$ & $\langle \dot{E}_{\rm cool}^{\rm cor} \rangle_t$ & $ z_T $ \\ 
           &  ($100$ orbits)&  (orbits)      & $(\rho_0 H_z^5 \Omega^3)$ & $(\dot{E}_{\rm turb})$ & $(\dot{E}_{\rm turb})$ & $(\rho_0 H_z^5 \Omega^3)$ & $(\dot{E}_{\rm turb})$ & $(\dot{E}_{\rm turb})$ & $(H_z$)          \\  
\hline
$\rm{NF} \: \: \: \: \mathcal{A} = 10 \: \: \: \beta_0 = 10^4$ & $1.5 \pm 0.4$ & $4.0 \pm 0.8$ & $5.8_{-1.0}^{+0.9}$ & $0.4_{-0.1}^{+0.2}$ & $0.6 \pm 0.1$ & $3.8_{-0.7}^{+0.6}$ & $0.10 \pm 0.04$ & $0.05 \pm 0.02$ & $1.6$ \\
$\rm{NF} \: \: \: \mathcal{A} = 10^2 \: \:  \beta_0 = 10^4$ & $3.4 \pm 0.8$ & $3.8 \pm 0.5$ & $4.8_{-0.5}^{+0.6}$ & $0.2 \pm 0.1$ & $0.8 \pm 0.1$ & $3.5 \pm 0.5$ & $0.06 \pm 0.04$ & $0.09 \pm 0.04$ & $2.2$ \\
$\rm{NF} \: \: \: \mathcal{A} = 10^3 \: \:  \beta_0 = 10^4$ & $6.2_{-2.1}^{+1.4}$ & $3.8 \pm 0.6$ & $4.4_{-0.6}^{+0.7}$ & $0.2 \pm 0.05$ & $0.9 \pm 0.2$ & $3.3 \pm 0.6$ & $0.05 \pm 0.04$ & $0.13 \pm 0.05$ & $2.9$ \\
$\rm{NF} \:    \rm{Isothermal}    \:     \beta_0 = 10^4$ & $5.7 \pm 1.9$ & $4.8 \pm 0.5$ & $3.7_{-0.5}^{+0.6}$ & $0.1 \pm 0.05$ & $0.9 \pm 0.05$ & $2.7_{-0.5}^{+0.6}$ & $0.06_{-0.05}^{+0.06}$ & --- & --- \\
\hline
$\rm{NF} \: \: \: \: \mathcal{A} = 10 \: \: \: \beta_0 = 10^3$ & $0.4 \pm 0.1$ & $0.7_{-0.3}^{+0.2}$ & $27.2_{-5.8}^{+6.1}$ & $0.4 \pm 0.1$ & $0.6 \pm 0.2$ & $18.0_{-5.3}^{+5.6}$ & $0.21 \pm 0.11$ & $0.17_{-0.07}^{+0.08}$ & $2.0$ \\
$\rm{NF} \: \: \: \mathcal{A} = 10^2 \: \:   \beta_0 = 10^3$ & $0.6 \pm 0.2$ & $0.8_{-0.4}^{+0.3}$ & $24.3_{-4.1}^{+4.3}$ & $0.3 \pm 0.1$ & $0.7 \pm 0.1$ & $16.9_{-4.5}^{+4.4}$ & $0.26 \pm 0.11$ & $0.26 \pm 0.09$ & $3.1$ \\
$\rm{NF} \: \: \: \mathcal{A} = 10^3 \: \:  \beta_0 = 10^3$ & $0.6 \pm 0.2$ & $0.6_{-0.2}^{+0.1}$ & $25.1 \pm 4.4$ & $0.2 \pm 0.1$ & $0.7 \pm 0.1$ & $16.3_{-4.4}^{+4.7}$ & $0.16_{-0.09}^{+0.08}$ & $0.22 \pm 0.06$ & $4.0$ \\
$\rm{NF} \:    \rm{Isothermal}    \:     \beta_0 = 10^3$ & $0.7_{-0.3}^{+0.2}$ & $0.7 \pm 0.2$ & $21.5_{-4.1}^{+4.2}$ & $0.2 \pm 0.1$ & $0.8 \pm 0.1$ & $15.3_{-4.5}^{+4.2}$ & $0.20 \pm 0.10$ & --- & --- \\
\hline
$\rm{ZNF} \: \: \mathcal{A} = 10 \: \: \: \: \: \beta_0 = 10^2$  & $14_{-3}^{+4}$ & $8.4_{-1.8}^{+1.5}$ & $2.1 \pm 0.4$ & $0.08 \pm 0.04$ & $1.0 \pm 0.2$ & $1.8_{-0.4}^{+0.3}$ & $0.03 \pm 0.03$ & $0.07 \pm 0.01$ & $1.6$ \\
$\rm{ZNF} \: \: \mathcal{A} = 10^2 \: \: \:\beta_0 = 10^2$  & $34 \pm 14$ & $8.4_{-1.7}^{+1.3}$ & $2.1_{-0.5}^{+0.4}$ &  $0.03 \pm 0.02$ & $1.0 \pm 0.2$ & $1.9 \pm 0.4$ & $0.05_{-0.03}^{+0.04}$ & $0.11 \pm 0.03$ & $2.3$ \\
$\rm{ZNF} \: \: \mathcal{A} = 10^3 \:\:\: \beta_0 = 10^2$  & $149_{-70}^{+66}$ & $10.8_{-1.7}^{+1.6}$ & $1.4 \pm 0.2$ & $0.01 \pm 0.01$ & $1.1 \pm 0.2$ & $1.3 \pm 0.2$ & $0.04 \pm 0.03$ & $0.11_{-0.03}^{+0.02}$ & $2.8$ \\
$\rm{ZNF} \:    \rm{Isothermal}    \:     \beta_0 = 10^2$  & $851_{-614}^{+607}$ & $16.4_{-2.5}^{+2.4}$ & $0.9 \pm 0.2$ & $0.004 \pm 0.004$ & $0.996 \pm 0.004$ & $0.8 \pm 0.1$ & $0.03 \pm 0.03$ & --- & --- \\
\hline
$\rm{ZNF} \: \: \: \: \mathcal{A} = 10 \: \: \: \beta_0 = 10$   & $15 \pm 3$ & $9.9_{-1.5}^{+1.4}$ & $1.6 \pm 0.3$ & $0.09 \pm 0.03$ & $1.0 \pm 0.2$ & $1.4_{-0.2}^{+0.3}$ & $0.03 \pm 0.03$ & $0.06 \pm 0.01$ & $1.6$ \\
$\rm{ZNF} \: \: \mathcal{A} = 10^2 \: \: \:  \beta_0 = 10$   & $33 \pm 12$ & $6.7_{-1.0}^{+1.1}$ & $2.3 \pm 0.4$ & $0.03 \pm 0.02$ & $1.0 \pm 0.2$ & $2.0 \pm 0.3$ & $0.05 \pm 0.03$ & $0.10 \pm 0.02$ & $2.3$ \\
$\rm{ZNF} \: \: \mathcal{A} = 10^3 \: \: \: \beta_0 = 10$   & $114 \pm 48$ & $9.3 \pm 2.0$ & $1.8 \pm 0.4$ & $0.02 \pm 0.01$ & $1.1 \pm 0.2$ & $1.6_{-0.4}^{+0.3}$ & $0.05 \pm 0.03$ & $0.11_{-0.03}^{+0.04}$ & $2.8$ \\
$\rm{ZNF} \:    \rm{Isothermal}    \:       \beta_0 = 10$   & $827_{-643}^{+595}$ & $13.1_{-2.1}^{+2.0}$ & $1.0 \pm 0.3$ & $0.005 \pm 0.005$ & $0.995 \pm 0.005$ & $0.9_{-0.2}^{+0.3}$ & $0.03_{-0.03}^{+0.02}$ & --- & --- \\
\hline
\end{tabular}}
%\end{centre}
\caption{Summary of key diagnostics for all 16~simulations. Quantities are quoted with $1 \sigma$ error bars, where the errors are computed from the 16th and 84th percentiles of the respective time series from 30 to 100 orbits. The box-averaged turbulent injection power $(\dot{E}_{\rm turb})$ and the power injected into the disc ($\dot{E}_{\rm turb}^{\rm disc}$) are measured in code units, while the box-averaged energy-outflow rate ($\dot{E}_{\rm wind}$), cooling rate ($\dot{E}_{\rm cool}$), Poynting flux into the corona ($\dot{E}_{\rm Poyt}^{\rm cor}$), and cooling within the corona ($\dot{E}_{\rm cool}^{\rm cor}$) are measured relative to $\dot{E}_{\rm turb}$. Horizontally averaged temperature first rises above the midplane temperature $T_0$ at height $z_T$. Isothermal simulations do not have any cooling. Thus, for isothermal simulations, the box-averaged cooling rate in the above table is defined as $\langle \dot{E}_{\rm cool} \rangle_t = \langle \dot{E}_{\rm turb} \rangle_t - \langle \dot{E}_{\rm wind} \rangle_t$.
} 
\label{table:simulations} 
\end{table*}

\subsection{Numerical details}
\label{sec:numerics}

The solutions are advanced in time using second-order-accurate van Leer time integration \citep[\texttt{vl2};][]{vanLeer1979}. Spatial integration is performed using the Harten--Lax--van Leer Discontinuity (\texttt{HLLD}) Riemann solver with second-order-accurate piecewise-linear-method (\texttt{PLM}) reconstruction. Details of the implementation of the stratified shearing box in \texttt{Athena++} can be found in \cite{Stone2010}.

Unlike most previous works, we do \textit{not} use orbital advection to advance our solution. The Courant condition is set by the \alf speed in the corona, not the large shear velocity at the radial boundaries. Thus, orbital advection is unnecessary in these simulations.

Following \cite{Hawley1995}, the MRI is seeded by randomly distributed, spatially uncorrelated velocity and adiabatic pressure fluctuations with maximum amplitude $\delta \boldsymbol{v} = 5 \times 10^{-3} c_{\rm s}^{\rm adi}/\gamma^{1/2}$ and $\delta P/P = 0.025$, where the adiabatic sound speed is defined $c_{\rm s}^{\rm adi} = \sqrt{\gamma P/ \rho}$. Perturbations are confined to $|z| \leq 0.5 H_z$ initially, and adjusted so that the mean momentum of the perturbations is zero. 

Finally, the gravitational potential used in the simulations is not formally that given by Equation~\ref{eq:gravitational_potential}, but rather the smoothed potential implemented in \cite{Davis2010},
\begin{equation} \label{eq:Phi_smooth}
    \Phi_{\rm smooth} = \left( \left[ \left( \zeta \mp 1 \right)^2 + \zeta^2 \lambda^2 \right]^{1/2} \mp \zeta \right)^2 \frac{1}{2} \Omega^2 z^2,
\end{equation}
where $\zeta = z_{\rm max}/z$ and $\lambda = 0.1 H_z/z_{\rm max}$. Here, minus and plus signs apply to regions above and below the midplane, respectively. For any computation involving $\Phi$, we use $\Phi_{\rm smooth}$ in place of $\Phi$.

\subsection{Simulations and diagnostics}
\label{sec:diagnostics}

We scan over three values of $\mathcal{A} \in \{10, 10^2, 10^3\}$ and four total field configurations: weak NF ($\beta_0 = 10^4$), moderate NF ($\beta_0 = 10^3$), weak ZNF ($\beta_0 = 10^2$), and moderate ZNF ($\beta_0 = 10$). Isothermal runs with each of the field configurations are provided for comparison. All simulations studied in this work, along with relevant timescales and box-integrated energy diagnostics, are given in Table~\ref{table:simulations}. 

The mass outflow rate is quantified via the wind depletion time,
\begin{equation} \label{eq:tdep}
    t_{\rm dep} \equiv \frac{M_{\rm box}}{\dot{M}_{\rm wind}},
\end{equation} 
which is compared to the thermal timescale of the disc,
\begin{equation} \label{eq:tthermal}
    t_{\rm thm} \equiv \frac{P_0}{\langle \mathcal{T}_{r \varphi} \rangle_{r \varphi t} (z = 0)} \frac{1}{\Omega} = \frac{1}{\alpha_{\rm mid} \Omega}.
\end{equation}
Here, we have introduced the \cite{Shakura1973} `$\alpha$' parameter, which we define in terms of the midplane pressure $P_0$ such that $\alpha \equiv \langle \mathcal{T}_{r \varphi}/P_0 \rangle_{r \varphi t}$; we define $\alpha_{\rm mid} \equiv \alpha (|z| < H_z/2)$. Following \cite{Stone1996}, the time average of a quantity $\mathcal{Q}$ is defined by
\begin{equation}
    \langle \mathcal{Q} \rangle_t \equiv \frac{1}{70 \: \rm{Orbits}} \int_{30 \: \rm{Orbits}}^{100 \: \rm{Orbits}} \mathcal{Q} (t, r, \varphi, z) \: \rmd t.
\end{equation}
Similarly, the horizontal and vertical averages are, respectively,
\begin{equation} \label{eq:horizontal_average}
    \langle \mathcal{Q} \rangle_{r \varphi} \equiv \frac{1}{L_x L_y} \int_{-L_y/2}^{+L_y/2} \int_{-L_x/2}^{+L_x/2} \mathcal{Q} (t, r, \varphi, z) \: \rmd x \rmd y,
\end{equation}
\begin{equation} \label{eq:z_average}
    \langle \mathcal{Q} \rangle_{z} \equiv \frac{1}{L_z} \int_{-L_z/2}^{+L_z/2} \mathcal{Q} (t, r, \varphi, z) \: \rmd z.
\end{equation}
The notation $\langle \mathcal{T}_{r \varphi} \rangle_{r \varphi z t}$ is then shorthand for $\langle \langle \langle \mathcal{T}_{r \varphi} \rangle_{r \varphi} \rangle_{z} \rangle_t$.

For NF configurations, our 100-orbit run time is ${\gtrsim} 25$ thermal times; for the ZNF configurations, the run time is ${\gtrsim} 10$ thermal times. The midplane $\alpha$ as well as box-averaged cooling and mass outflow rates are saturated after 30 orbits in all simulations, with all but the isothermal ZNF simulations attaining saturation around 20 orbits.

\begin{figure*}
\hbox{
\includegraphics[width=1.0\textwidth]{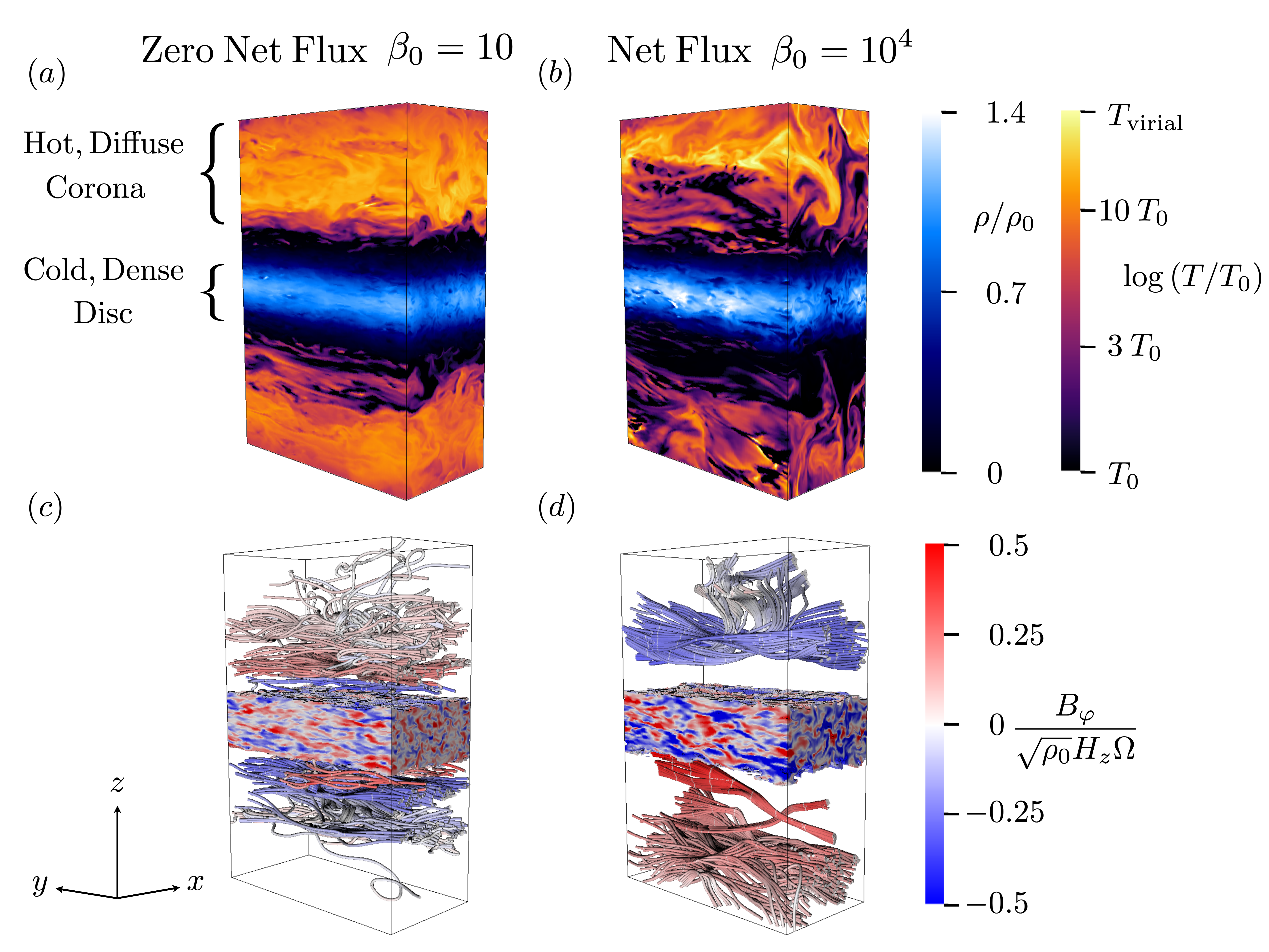}
}
\caption{Top row: Volume renderings of temperature ($T$) and density ($\rho$) for the zero net flux (ZNF) $\mathcal{A} = 10$ $\beta_0 = 10$ (panel a) and net flux (NF) $\mathcal{A} = 10$ $\beta_0 = 10^4$ (panel b) simulations at times $t = 61.8$ orbits and $t = 64.7$ orbits, respectively. Bottom row: magnetic-field line renderings, with color indicating toroidal field ($B_{\varphi}$) for the ZNF $\mathcal{A} = 10$ $\beta_0 = 10$ (panel c) and NF $\mathcal{A} = 10$ $\beta_0 = 10^4$ (panel d) simulations. Because toroidal field reversals are so frequent in the turbulent disc, a volume rendering (rather than streamlines) of the toroidal field is used around the midplane. In panels (a) and (b), temperature is only shown where $T > 1.2 T_0$, with $T_0$ as the midplane temperature. Density is shown where temperature is below this threshold to emphasize the structure of the thin disc. 
}
\label{fig:volume_renderings}
\end{figure*}

When examining outflows, it is useful to introduce the energy outflow rate $\dot{E}_{\rm wind}$ measured at the vertical boundaries $(|z| = 6 H_z)$,
\begin{equation} \label{eq:Edot_wind}
    \dot{E}_{\rm wind} = \oiint \left( \rho u_z \mathcal{B} \right) \left( |z| = 6 H_z \right) \: \rmd x \rmd y,
\end{equation}
where for an ideal EOS the Bernoulli parameter $\mathcal{B}$ is 
\begin{equation} \label{eq:ideal_Bernoulli}
    \mathcal{B} \equiv \frac{1}{2} |\boldsymbol{u}|^2 + \frac{\gamma}{\gamma -1} \frac{P}{\rho} + \Phi - \frac{3}{2} \Omega^2 x^2 + \frac{1}{\rho u_z} \mathcal{S}_z.
\end{equation}
For an isothermal EOS \citep{Bai2013_NF},
\begin{equation} \label{eq:isothermal_Bernoulli}
    \mathcal{B} \equiv \frac{1}{2} |\boldsymbol{u}|^2 - c_{\rm s0}^2 \log{\left( \rho/\rho_0 \right)} + \Phi - \frac{3}{2} \Omega^2 x^2 + \frac{1}{\rho u_z} \mathcal{S}_z.
\end{equation}
Here, the Poynting flux is defined by
\begin{equation} \label{eq:Poynting_flux}
    \boldsymbol{\mathcal{S}} \equiv \boldsymbol{B} \btimes \boldsymbol{u} \btimes \boldsymbol{B} = \frac{|\boldsymbol{B}|^2}{2} \boldsymbol{u} - \boldsymbol{B} \left( \boldsymbol{B} \bcdot \boldsymbol{u} \right),
\end{equation}
and the Poynting flux in the vertical direction is $\mathcal{S}_z \equiv \boldsymbol{S} \bcdot \hat{\bb{z}}$. Note that the \textit{total} velocity $\boldsymbol{u}$, rather than the \textit{turbulent} velocity $\boldsymbol{v}$, enters the expressions for $\mathcal{B}$, as is the case in the total energy equation (\ref{eq:energy}).

We define the corona as the region above the $|z| = 2 H_z$ surface. Because we do not include self-consistent radiation transport, this definition is entirely arbitrary. In a self-consistent simulation, the corona would be defined by the surface at which $\tau_{\rm es} = 1$, i.e., the point above which the flow transitions from optically thick to thin. We shall see, however, that defining the beginning of the corona as $|z| = 2 H_z$ roughly captures the point where the temperature increases significantly and the disc atmosphere becomes magnetically dominated. Thus, the chosen coronal height is reasonably physically motivated for our idealized problem. We refer to the height where the horizontally averaged temperature, $\langle T \rangle_{r \varphi t}$, first rises above $T_0$ as $z_T$, and we list $z_T$ for each simulation in Table~\ref{table:simulations}. With our definition of the corona in mind, the Poynting flux into the corona is given by
\begin{equation}
    \dot{E}_{\rm Poyt}^{\rm cor} = \oiint \boldsymbol{\mathcal{S}} \bcdot \hat{\bb{z}} \left( |z| = 2 H_z \right) \: \rmd x \rmd y,
\end{equation}
where the surface integral is taken over both planes at $z = \pm 2 H_z$.

\begin{figure*}
\hbox{
\includegraphics[width=1.0\textwidth]{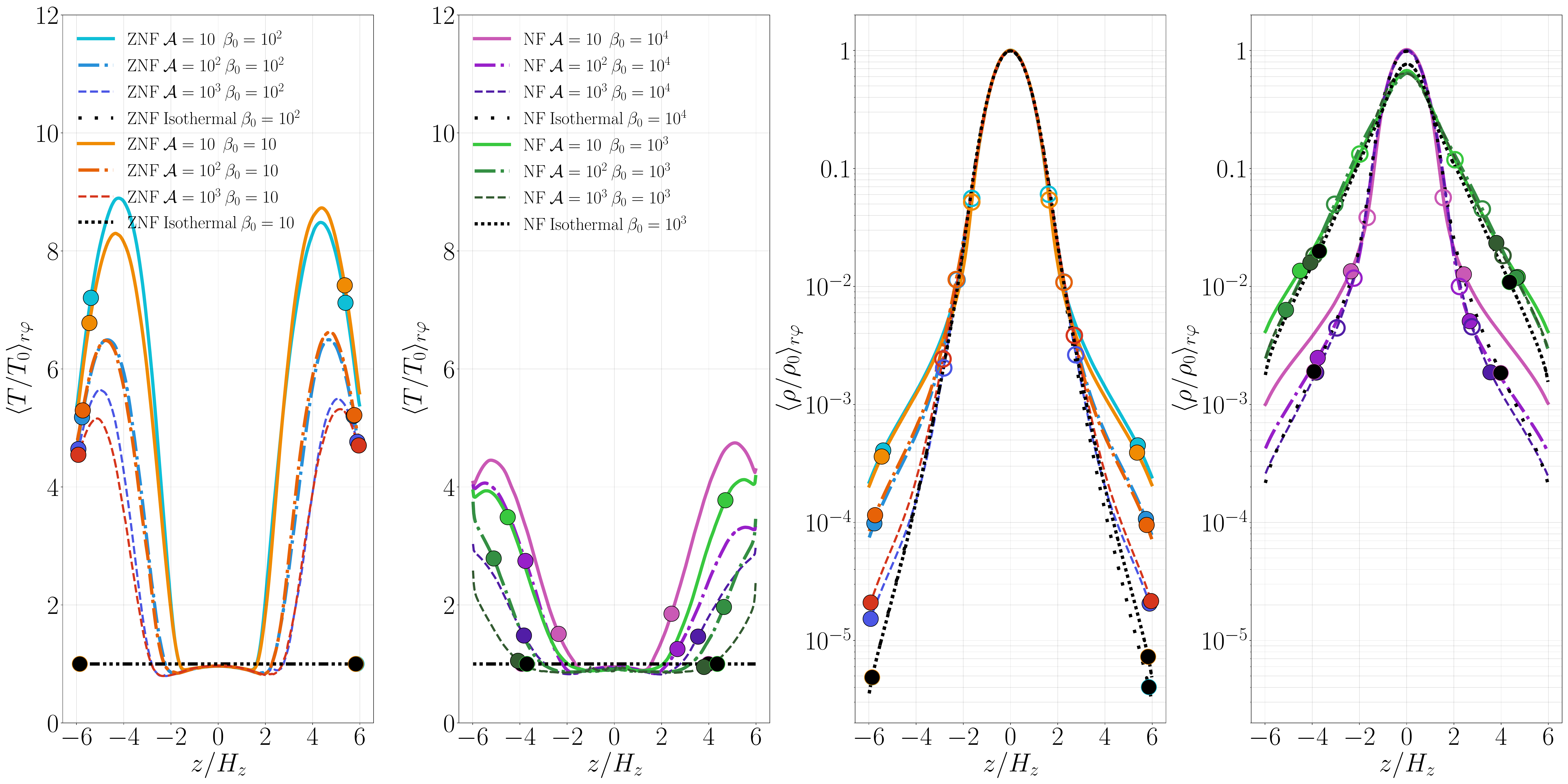}
}
\caption{Left: Horizontally-averaged temperature ($T$) profiles for zero net-flux (ZNF) and net flux (NF) simulations from 30--100 orbits. Right: Horizontally-averaged density ($\rho$) profiles for ZNF and NF simulations for the same time frame. Filled circles indicate slow magnetosonic points ($z_{\rm s}$) while open circles show where $T$ rises above $T_0$ (where $z = z_T$). All ideal equation of state simulations show temperature inversions with a hotter corona surrounding a colder disc.
}
\label{fig:temperature_density}
\end{figure*}

\section{Results: flow structure}
\label{sec:flow_structure}

Figure~\ref{fig:volume_renderings} demonstrates the key features of our simulations. By suspending the isothermality assumption, the simulations allow the development of temperature inversions: a high temperature `corona' surrounding a cold `disc' (Figures~\ref{fig:volume_renderings}a and \ref{fig:volume_renderings}b; \S\ref{sec:temperature_inversions}). Intermittent heating and Coulomb cooling form a `multi-phase corona' with broadened density and temperature distributions dependent upon the interplay of outflows (\S\ref{sec:thermally_driven_winds}) and cooling (\S\ref{sec:multiphase_corona}). 

\newpage

While conduction is not included in these simulations, temperature inversions combined with field lines extending between the disc and corona would allow a field-aligned conductive coupling between the corona and disc. The dominantly toroidal field geometrically suppresses the heat flux relative to what would be expected purely from the vertical temperature gradient. We provide estimates for the magnitude of this suppression in \S\ref{sec:conduction}. 

Field lines extending out from the corona (Figure~\ref{fig:volume_renderings}d) enable NF runs to launch magnetically driven outflows, which rapidly remove mass and angular momentum from the accretion disc, potentially evacuating a global disc of material (\S\ref{sec:wind_accretion}) and aiding evaporation of the radiatively efficient disc into a radiatively inefficient accretion flow (RIAF). In ZNF simulations, MRI turbulence, outflows, and buoyancy form complex magnetic-field configurations composed of twisted flux ropes and loop structures (Figure~\ref{fig:volume_renderings}c). Magnetic energy transport and dissipation within the corona are affected by thermodynamics as well. In \S\ref{sec:Poynting_cooling}, we quantity the Poynting fluxes through the $|z| = 2 H_z$ surface as well as the amount of cooling in the corona.

\subsection{Temperature inversions, magnetic structure, and accretion}
\label{sec:temperature_inversions}

Figure~\ref{fig:temperature_density} shows horizontally averaged density and temperature profiles for all simulations. Temperature inversions occur in all ideal EOS simulations for all values of $\mathcal{A}$ studied in this work. 

The value of $\mathcal{A}$ has two effects: (1) stronger Coulomb coupling (larger $\mathcal{A}$) implies a smaller horizontally averaged temperature in the corona above $|z| = 2 H_z$, and (2) stronger Coulomb coupling implies that temperatures rise above the midplane temperature $T_0$ at higher $|z|$. Coronal plasma obeys an internal energy equation,
\begin{equation} \label{eq:internal_energy}
    \rho \D{t}{} \frac{T}{\gamma - 1} = - \underbrace{P \left( \grad \bcdot \boldsymbol{u} \right)}_{\rm compression} + \underbrace{Q_{\rm heat}}_{\rm heating} - \underbrace{Q_{\rm cool}^{-}}_{\rm cooling} + \underbrace{ \frac{1}{\gamma - 1} \dot{\rho}_{\rm src} T}_{\rm injection},
\end{equation}
where we have introduced the Lagrangian derivative in the fluid frame, $\rmd /\rmd t \equiv ( \partial/\partial t + \boldsymbol{u} \bcdot \grad)$. Temperature increases when heating $Q_{\rm heat}$ outpaces cooling $Q_{\rm cool}^{-}$ and decompression $P (\grad \bcdot \boldsymbol{u})$. Thus, temperature inversions are a consequence of both the rate of dissipation $Q_{\rm heat}$ and the density structure, which sets $Q_{\rm cool}^{-}$. 

Initially subthermal ZNF fields, such as those in the $\beta_0 = 10$ and $\beta_0 = 10^2$ models, achieve similar saturated field configurations. The cooling rate, as controlled by $\mathcal{A}$, is the primary determinant of the temperature structure. Figure~\ref{fig:temperature_density} displays this property of our solutions: ZNF simulations with the same $\mathcal{A}$ and different initial field strengths show similar temperature profiles.

In the ZNF simulations, density drops off steeply with height (Figure~\ref{fig:temperature_density}). While the initial Gaussian profile is maintained in all simulations until a height of ${\approx}2 H_z$, ideal EOS runs show significantly enhanced density above $|z| \approx 3 H_z$ relative to the isothermal reference runs. When cooling is sufficiently weak ($\mathcal{A} \lesssim 10^2$), densities at fixed height in ideal EOS runs can be larger than the isothermal reference cases by an order of magnitude or more. 

Because our two-temperature cooling function is ${\propto} \rho^2$ (see Equation~\ref{eq:cooling_function}), the rate of cooling $Q_{\rm cool}^{-}$ decreases rapidly with height, allowing the coronal temperature to increase. Indeed, the existence of the temperature inversions implies that $Q_{\rm heat} > Q_{\rm cool}^{-}$ above heights $|z| \approx 2 H_z$. To maintain the measured temperature of $T =$~8.5~$T_0$ in the $\mathcal{A} = 10$ runs just above $|z| = 4 H_z$, the heating rate need only be $Q_{\rm heat} \approx (10) \times (10^{-3})^2 \times 8.5 \rho_0 T_0 \Omega = 4.3 \times 10^{-5} \rho_0 H_z^2 \Omega^3$ . Comparing this heating rate to the box-averaged turbulent injection rate $\dot{E}_{\rm turb}/L_x L_y L_z$ from Table~\ref{table:simulations} in the $\mathcal{A} = 10$, $\beta_0 = 10$ simulation, we find that only ${\approx} 1$ per cent of the injected energy must be dissipated as heat above $|z| = 4 H_z$ to sustain our measured coronal temperatures. Profiles of the cooling rate $Q_{\rm cool}^{-} (z)$ are provided in \S\ref{sec:Poynting_cooling}.

\begin{figure*}
\hbox{
\includegraphics[width=1.0\textwidth]{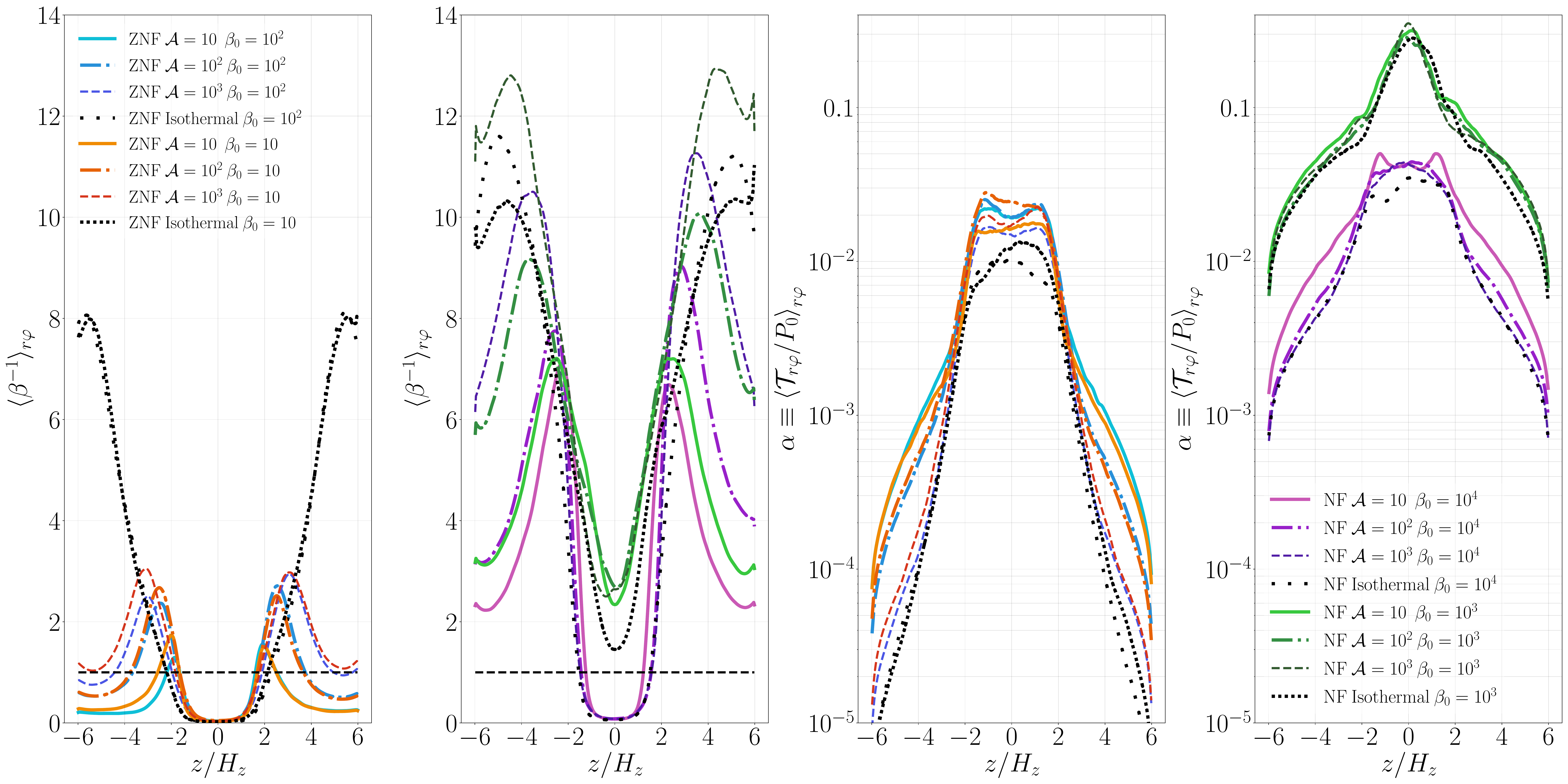}
}
\caption{Left: Horizontally-averaged $\beta^{-1}$ parameter (Equation~\ref{eq:beta_inv}) from 30--100 orbits. A line is drawn at $\langle \beta^{-1} \rangle_{r \varphi} = 1$. Right:  Horizontally-averaged turbulent $r-\varphi$ stress (i.e. $\alpha$ parameter) from 30--100 orbits. Magnetizations $(\langle \beta^{-1} \rangle_{r \varphi})$ are modest in the zero net-flux (ZNF) simulations, and $\langle \beta^{-1} \rangle_{r \varphi}$ decreases with weaker cooling (decreasing $\mathcal{A}$), independent of field configuration. For the weak and moderate ZNF fields explored in this work, the saturated $\alpha$ is nearly independent of initial field strength and $\mathcal{A}$. NF simulations display higher accretion rates for larger initial vertical magnetic fluxes, but the midplane $\alpha$ is still nearly independent of $\mathcal{A}$. Accretion in the surface layers of the disc is enhanced in weakly cooled ($\mathcal{A} = 10$) runs.
}
\label{fig:magnetization_alpha}
\end{figure*}

To assess the effect of temperature inversions on the magnetic structure of the corona, we introduce the magnetization parameter $\beta^{-1}$, which is the inverse of the plasma $\beta$. We define the horizontal average of $\beta^{-1}$ as
\begin{equation} \label{eq:beta_inv}
    \langle \beta^{-1} \rangle_{r \varphi} \equiv \frac{\langle B^2/2 \rangle_{r \varphi t}}{\langle \rho T \rangle_{r \varphi t}}.
\end{equation}
Here, we have averaged the magnetic energy density (numerator) and thermal pressure (denominator) separately to avoid over-weighting cells with enormous values of $\beta^{-1}$. %which are numerical artifacts.

Figure~\ref{fig:magnetization_alpha} shows profiles of $\langle \beta^{-1} \rangle_{r \varphi}$. Our simulations all form hot, magnetized ($\langle \beta^{-1} \rangle_{r \varphi} > 1$) coronae above $|z| = 2 H_z$. However, compared to isothermal simulations, the two-temperature ideal EOS simulations display modest peak $\beta^{-1}$ values. Weaker cooling (smaller $\mathcal{A}$) results in a weaker magnetization $\beta^{-1}$ above $|z| > 2 H_z$, independent of initial field configuration.

This decrease in magnetization with decreasing $\mathcal{A}$ is driven primarily by the increase in temperature in the ZNF ideal EOS simulations. The exception is the ZNF $\mathcal{A} = 10$ simulations, which exhibit an order of magnitude density enhancement over the isothermal reference calculations. Increased coronal temperatures and densities imply that weakly cooled ($\mathcal{A} \lesssim 10^2$) ZNF simulations are weakly magnetized ($\langle \beta^{-1} \rangle_{r \varphi} < 1$) above $|z| = 4 H_z$. However, $\langle \beta^{-1} \rangle_{r \varphi}$ profiles are similar through $|z| = 3 H_z$ in all ZNF simulations.

In NF simulations, density structure is influenced more by the initial vertical magnetic flux than by the cooling rate (Figure~\ref{fig:temperature_density}). Density profiles in the weak-field ($\beta_0 = 10^4$) cases follow an approximate Gaussian distribution within $|z| \approx 2 H_z$ of the midplane before flattening to a power law with $\rmd\ln\rho/\rmd\ln|z| \approx -4$. These profiles are largely independent of $\mathcal{A}$, with the cooling strength setting the height at which the profile transitions from a Gaussian to a power law. 

Density profiles in moderate-field ($\beta_0 = 10^3$) NF situations differ substantially from the weaker, NF $\beta_0 =10^4$ simulations. Instead of a Gaussian distribution, the time-averaged profiles settle into an approximately exponential (in $-|z|$) distribution \citep{Bai2013_NF}. Densities in the corona are nearly an order of magnitude higher in the moderate-field cases compared to those of the weak-field cases for all but the $\mathcal{A} = 10$ simulations. While there is some dependence on thermodynamics, with weaker cooling enhancing density at fixed height, the differences are nowhere as pronounced as in the weaker NF runs. The $\mathcal{A} = 10$ run achieves approximately twice the horizontally averaged density of the isothermal reference run at $|z| = 4H_z$.

High coronal densities have a significant effect on the temperatures attained above $|z| = 2 H_z$. While all ideal EOS NF simulations show temperature inversions, the maximum horizontally averaged temperatures are lower than in any ZNF simulations. The temperature of $\langle T \rangle_{r \varphi t} \approx 4.3 T_0$ at $|z| \approx 5 H_z$ in the NF $\mathcal{A} = 10$, $\beta_0 = 10^4$ simulation can be sustained by a heating rate of $Q_{\rm heat} \approx 1.7 \times 10^{-4} \rho_0 H_z^2 \Omega^3$, again, ${\approx} 1\%$ of the injected energy. Note that this estimate for the heating efficiency comes from higher up in the corona than the comparable measurement in the ZNF $\mathcal{A} = 10$, $\beta_0 = 10$ case ($|z| = 5 H_z$ instead of $4 H_z$), implying that heating in NF simulations is more efficient than in ZNF runs (see \S\ref{sec:Poynting_cooling}).

Despite increased densities in their coronae, NF simulations are always strongly magnetized above $|z| = 2 H_z$. As in the ZNF cases, magnetization $\langle \beta^{-1} \rangle_{r \varphi}$ decreases with decreasing $\mathcal{A}$ in NF simulations, and magnetization profiles (Figure~\ref{fig:magnetization_alpha}) are quite similar from $2 \lesssim |z|/H_z \lesssim 3$, independent of thermodynamics or initial $\beta_0$. Since density is largely independent of $\mathcal{A}$ in NF simulations, particularly for the moderate NF cases, magnetization is primarily determined by variations in temperature, not density. Unlike the ZNF or weak NF cases, moderate NF simulations are magnetically dominated at \textit{all} heights, including at the midplane, and $\langle \beta^{-1} \rangle_{r \varphi}$ at the midplane is actually \textit{larger} in the ideal EOS runs relative to isothermal runs.

The right panels of Figure~\ref{fig:magnetization_alpha} display the $\alpha$ parameters measured in our disc simulations. As expected, $\alpha$ increases with increasing NF \citep{Salvesen2016_NF}. In ZNF simulations, $\alpha$ is largely independent of initial field strength; however, there is some dependence on thermodynamics in the corona above $|z| = 2 H_z$. For instance, at $|z| = 4 H_z$, the $\mathcal{A} = 10$ simulations display $\alpha$ parameters about an order of magnitude larger than isothermal simulations with the same initial field configuration. While thermal times, based on the midplane $\alpha$, are all similar across ZNF simulations (Table~\ref{table:simulations}), accretion proceeds more rapidly through our two-temperature coronae.

Enhanced accretion with weaker cooling (smaller $\mathcal{A}$) is evident in NF simulations as well, although primarily for weak ($\beta_0 = 10^4$) NF fields. Accretion is largely independent of thermodynamics for moderate ($\beta_0 = 10^3$) NF fields, but for weak fields, $\alpha$ in the $\mathcal{A} = 10$ simulation exceeds that in the isothermal reference by a factor of ${\approx}2.5$ at $|z| = 4 H_z$. We find weak signatures of magnetically elevated accretion \citep{Begelman2017} in this $\mathcal{A} = 10$ weak NF run, with $\alpha$ increased by ${\approx}25$ per cent in a narrow layer around $|z| = H_z$. This increase becomes more apparent with taller boxes (see Appendix~\ref{sec:height_scan}).

\begin{figure}
\hbox{
\includegraphics[width=0.48\textwidth]{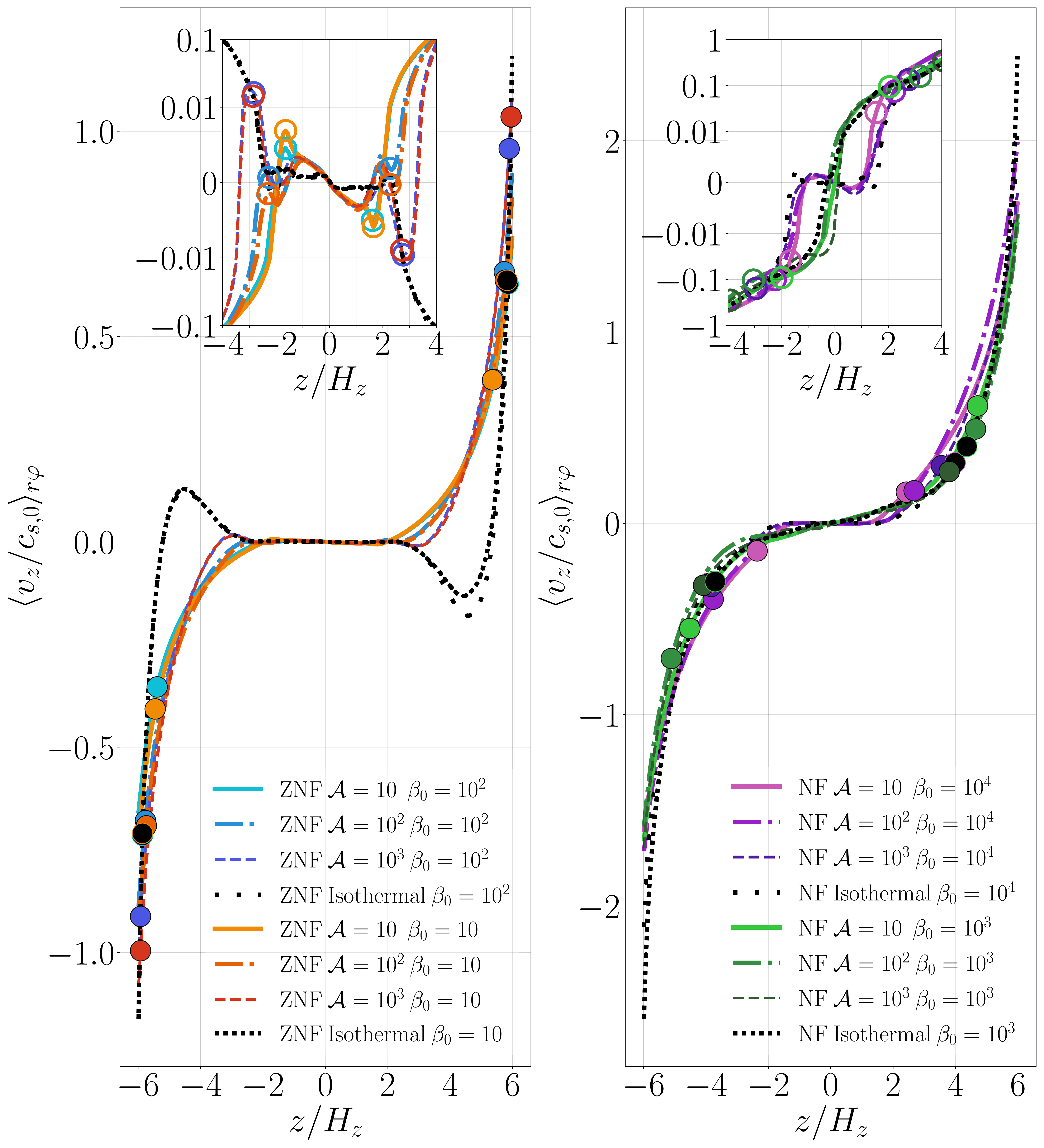}
}
\caption{Horizontally averaged profiles of vertical velocity ($v_z$)  for zero net-flux (ZNF) simulations (left) and net-flux (NF) simulations (right), averaged from 30--100 orbits. Filled circles indicate slow magnetosonic points while open circles show $z_T$, where the horizontally averaged temperature first rises above the midplane temperature. All simulations show an outflow characteristic of a wind. In NF simulations, this wind is magnetocentrifugally driven (see text); however, ZNF runs are likely thermally driven, as indicated by the correlation of $z_T$ with inflection points in the $v_z$ profiles (see inset plots).
}
\label{fig:vz_profiles}
\end{figure}

\subsection{Thermally vs.~magnetocentrifugally driven outflows}
\label{sec:thermally_driven_winds}

Figure~\ref{fig:vz_profiles} shows horizontally averaged profiles of vertical velocity ($\langle v_z \rangle_{r \varphi}$) for all simulations. At $|z| = 6 H_z$, all simulations exhibit an outflow. Independent of initial field strength, the outflow is characteristic of a wind---a steady acceleration with increasing height. 

A key question is whether winds in our simulations are primarily thermally driven, via gradients in gas thermal pressure \citep{Begelman1983}, or magnetocentrifugally driven, via rotation of open field lines extending from the disc \citep[Figure~\ref{fig:volume_renderings}d;][]{Blandford1982}. We find clear evidence that heating and therefore thermal driving accelerates winds in the ZNF simulations, with the most striking evidence visible in the inset plot of the left panel of Figure~\ref{fig:vz_profiles}. The heights of $z_T$ (shown as open circles) correlate with inflection points in the $\langle v_z \rangle_{r \varphi t} (z)$ curve in all ideal EOS simulations. At these points, the flow is accelerating and transitioning into an outflow. 

Similar correlations between $z_T$ and inflection points in $\langle v_z \rangle_{r \varphi}$ are not evident in NF simulations. Instead, the outflow begins well below $|z| = z_T$, suggesting magnetic rather than thermal acceleration. Indeed, by comparing the thermal pressure gradient $\rmd P/\rmd z$ to the toroidal field pressure gradient $\rmd  (B_{\varphi}^2/2)/\rmd z$, we find that thermal driving is insignificant in all NF simulations. Between heights of $|z| = (2-5) H_z$ (the range in which the flows achieve a slow magnetosonic---hereafter, `sonic'---transition; see Figure~\ref{fig:vz_profiles}, right panel) thermal driving contributes between 5 and 30 per cent of the total pressure gradient. Weak-field simulations show greater thermal driving (30 vs. 10 per cent in the moderate field runs) near $|z| \approx 5H_z$, yet thermal driving is certainly subdominant.

The same comparison of thermal vs. toroidal magnetic pressure gradients in the ZNF runs indicates that, near $z_T$, thermal driving is again subdominant, contributing only ${\approx}20\%$ of the total vertical pressure gradient. However, thermal driving takes over above $3H_z$, $4 H_z$, and $5 H_z$ in the $\mathcal{A} = 10$, $\mathcal{A} = 10^2$, and $\mathcal{A} = 10^3$ models respectively. For the $\mathcal{A} = 10$ model in particular, once outflows achieve a sonic transition (filled points in left panel of Figure~\ref{fig:vz_profiles}), thermal pressure contributes ${\gtrsim} 80\%$ of the vertical pressure gradient, indicating that the outflows observed in the ZNF simulations are indeed thermally driven. 

Slow magnetosonic points (hereafter, `slow points') are located within the domain in all simulations. However, because the sonic transition occurs so close to the vertical boundary in ZNF simulations with $\mathcal{A} \geq 10^2$ and in isothermal runs, these strongly cooled flows may never actually attain a sonic transition within a height $|z| < 6 H_z$. In the $\mathcal{A} = 10$ ZNF simulations, there is sufficient separation between the slow point and the boundary to establish the presence of thermally driven supersonic outflows. Thus, mass outflow rates, which are used to compute the depletion times (Equation~\ref{eq:tdep}) in Table~\ref{table:simulations}, are likely robust in the ZNF $\mathcal{A} = 10$ and all NF simulations, but not necessarily in more strongly cooled ZNF runs. 

Indeed, our use of the term `winds' with regards to \textit{any} of the ZNF simulations requires caveats. Magnetic-field loops, as would be expected to form in these simulations (Figure~\ref{fig:volume_renderings}c), are not open at infinity, and so there may be no way for material to leave the system. Rather than true thermally driven outflows, we may instead be seeing the base of an `accretion disc fountain', similar to those discussed in the context of galaxy formation \citep{Hopkins2014}. In this picture, plasma is driven from the disc via thermal pressure, yet instead of escaping to infinity, the material cools and falls back down onto the disc, re-fueling the accretion flow in a cyclical manner. Resolving the fate of thermally driven outflows in ZNF fields may only be possible through global simulations with sufficient dynamic range to capture the fountain. We leave this question to future studies.

\begin{figure*}
\hbox{
\includegraphics[width=1.0\textwidth]{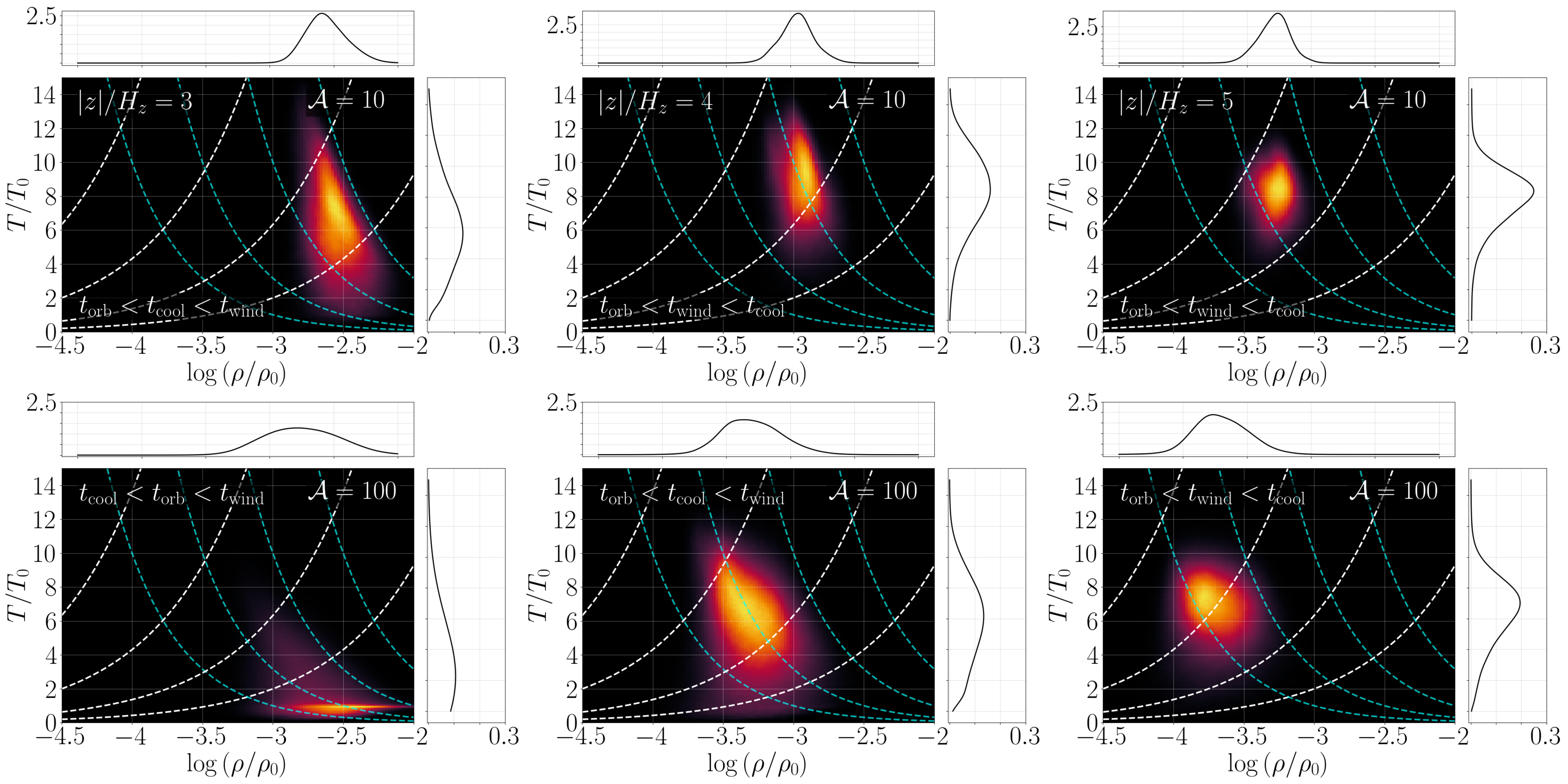}
}
\caption{Density-temperature distributions for all plasma in $\beta_0 = 10$ ZNF simulations at heights $|z| = 3 H_z$ (left column), $|z| = 4 H_z$ (centre column), and $|z| = 5 H_z$ (right column) from 30--100 orbits. The top row shows the $\mathcal{A} = 10$ simulation while the bottom displays data from the $\mathcal{A} = 100$ run. Cooling timescale $t_{\rm cool}$, wind outflow timescale $t_{\rm wind}$, and orbital timescale ($t_{\rm orb}$) are compared for each height/ simulation. Curves of constant entropy (white) and constant pressure (light blue) are drawn for reference. When $t_{\rm orb} < t_{\rm cool} < t_{\rm wind}$, extended density and temperature distributions form along lines of approximately constant thermal pressure, as would be expected for isobaric cooling in a thermal pressure dominated (rather than magnetic pressure dominated) plasma. 
}
\label{fig:thermal_instability_ZNF}
\end{figure*}

\subsection{Formation of a multiphase corona}
\label{sec:multiphase_corona}

Photons that are Compton upscattered to form the power-law signatures characteristic of AGN coronae traverse the density and temperature structure of the corona. Here, we assess if this journey is through a `clumpy' corona, with a broad range of densities and temperatures encountered, or if the lepton number density responsible for setting the optical depth $\tau_{\rm es}$ is relatively uniform at a given height. Using evidence provided by examining density and temperature distributions at different heights in our model coronae, we quantify the conditions that would allow a clumpy, multiphase corona to form. 

\begin{figure*}
\hbox{
\includegraphics[width=1.0\textwidth]{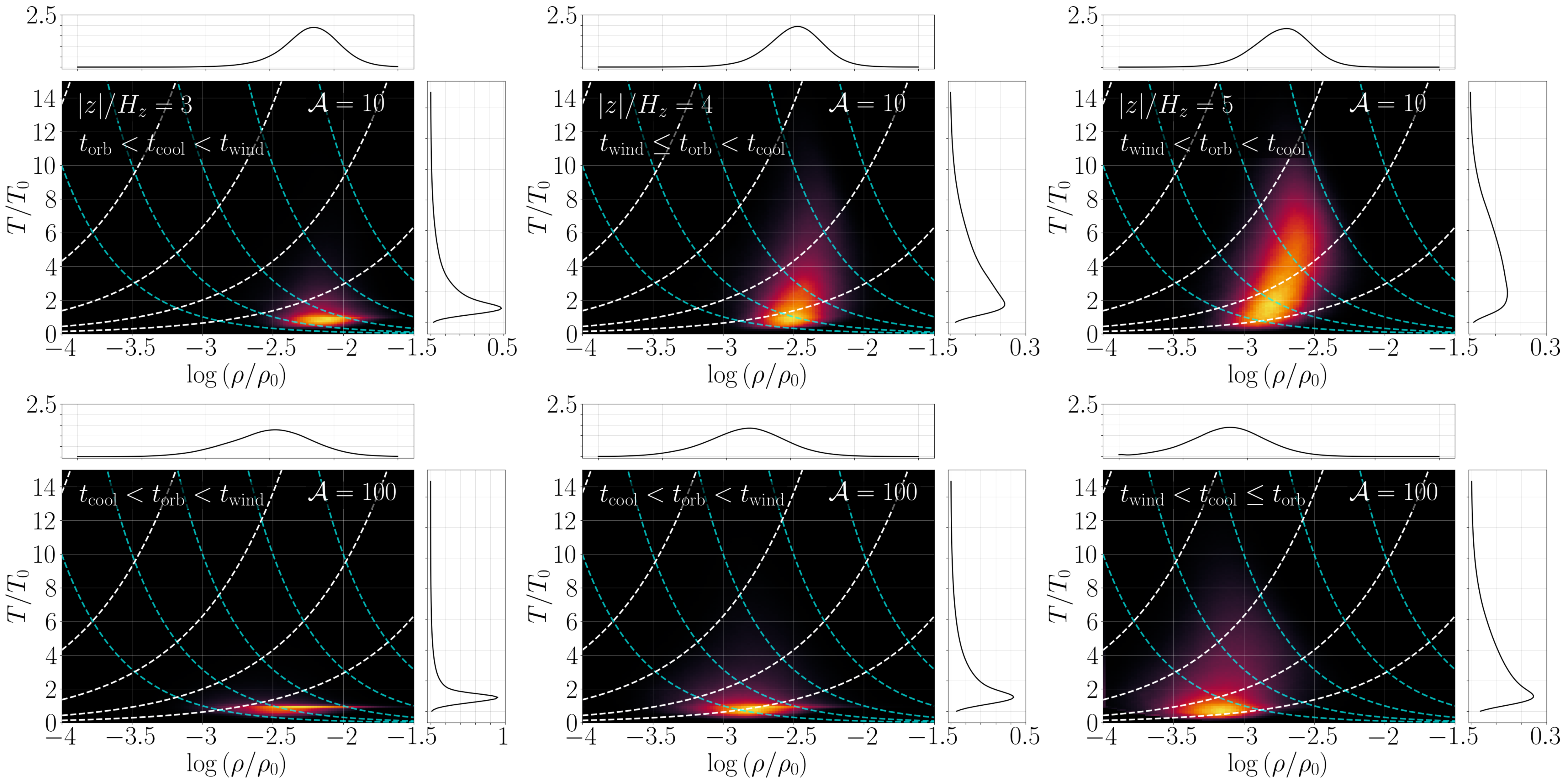}
}
\caption{Density-temperature distributions for all plasma in $\beta_0 = 10^4$ NF simulations at heights $|z| = 3 H_z$ (left column), $|z| = 4 H_z$ (centre column), and $|z| = 5 H_z$ (right column) from 30--100 orbits (similar to Figure~\ref{fig:thermal_instability_ZNF}). Curves of constant entropy (white) and constant pressure (light blue) are drawn for reference. Even when $t_{\rm wind} < t_{\rm orb} < t_{\rm cool}$, temperature increases with increasing density, crossing lines of constant pressure and entropy. Rather than isobaric cooling dictating thermal structure in these runs, the density-temperature correlation is a signature of fast advection of hot wind material through the corona. 
}
\label{fig:thermal_instability_NF}
\end{figure*}

Figure~\ref{fig:thermal_instability_ZNF} shows the density and temperature structure at three different heights in the ZNF $\mathcal{A} = 10$ and $\mathcal{A} = 10^2$ runs with an initial $\beta_0 = 10$ field. A similar view into the NF $\beta_0 = 10^4$ simulations can be found in Figure~\ref{fig:thermal_instability_NF}. For guidance, we over-plot curves of constant thermal pressure (light blue) and curves of constant entropy (white) onto the density and temperature ($\rho$--$T$) distributions in Figures~\ref{fig:thermal_instability_ZNF} and~\ref{fig:thermal_instability_NF}. The distributions at fixed height are quite different for simulations with different $\mathcal{A}$; even within the same simulation, the thermal structure of the corona can vary widely between $|z| = 3 H_z$ and $5 H_z$. 

To understand the origin of this thermal structure, we utilize similar techniques as have been developed to understand thermal instability \citep{Field1965} in the interstellar medium (ISM) and intracluster medium (ICM). Traditionally, criteria for thermal instability involve a comparison of the cooling time $t_{\rm cool}$ to the dynamical timescale of the system. In galaxy clusters, which maintain a quasi-thermal and quasi-hydrostatic equilibrium \citep{McNamara2007}, this dynamical time is taken to be the free-fall time \citep{McCourt2012, Sharma2012}. However, since our atmospheres are outflowing, there is a second relevant dynamical timescale \citep{Balbus1986, Waters2022} for comparison with $t_{\rm cool}$: the wind outflow time, $t_{\rm wind}$.

\begin{figure*}
\hbox{
\includegraphics[width=1.0\textwidth]{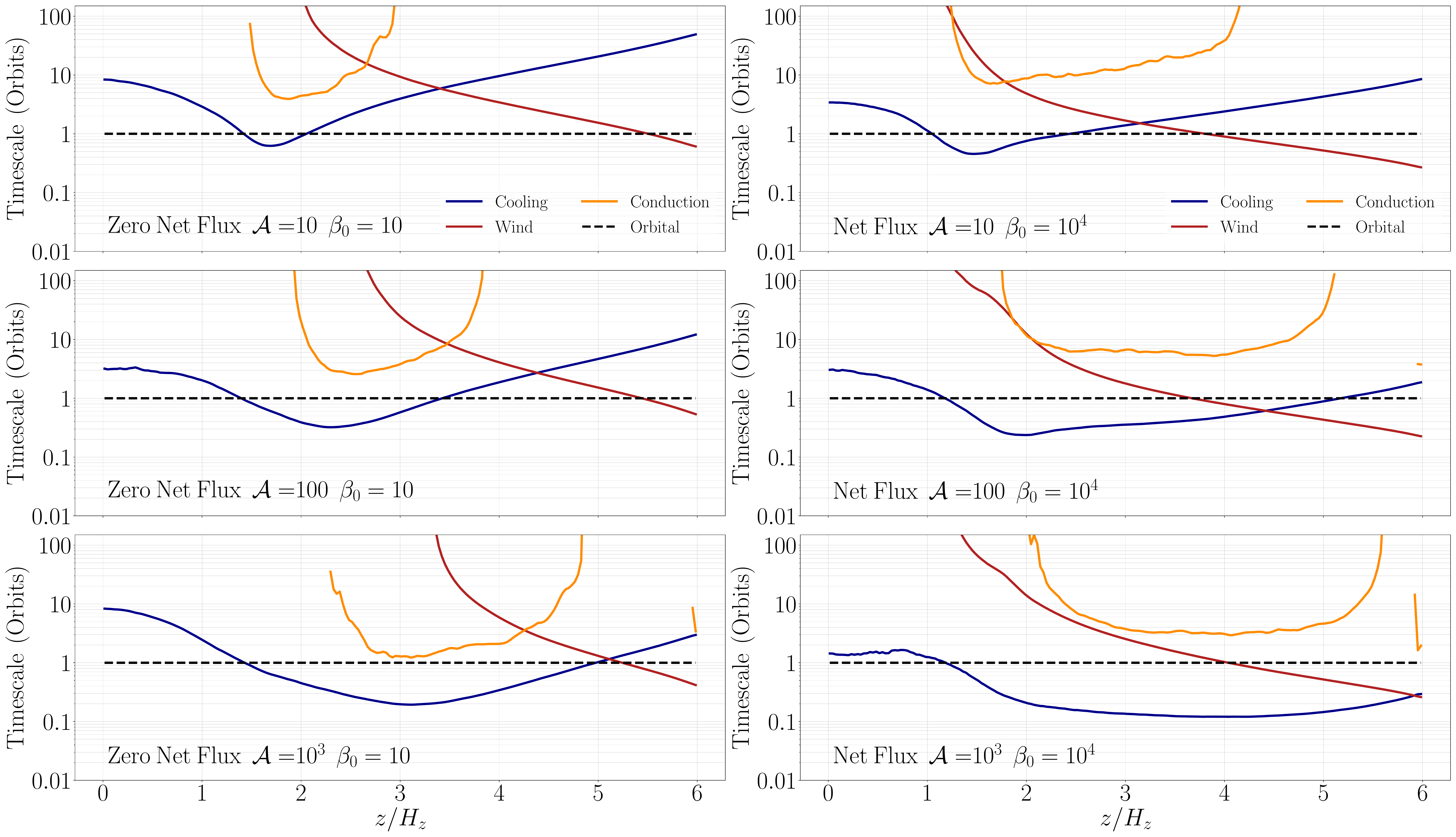}
}
\caption{Timescale profiles for zero net flux $\beta_0 = 10$ (left) and net flux $\beta_0 = 10^4$ (right) simulations from 30--100 orbits. We compare the wind outflow time $t_{\rm wind}$ (red; Equation~\ref{eq:t_wind}) to the measured cooling time $t_{\rm cool}$ (blue; Equation~\ref{eq:tcool}) and orbital timescale $t_{\rm orb}$ (black dashed). The conduction time $t_{\rm cond}$ (orange) is defined in Equation~\ref{eq:tcond}. A multiphase corona can form through cooling when $t_{\rm orb} < t_{\rm cool} < t_{\rm wind}$.
}
\label{fig:timescales}
\end{figure*}

We define the wind-outflow timescale $t_{\rm wind}$ as the time it takes for the wind to transport material upwards by a distance $2 H_z$ (i.e., from the disc midplane into the corona),
\begin{equation} \label{eq:t_wind}
    t_{\rm wind} (z) \equiv \frac{2 H_z}{|\langle v_z \rangle_{r \varphi t}|}.
\end{equation}
The cooling time is defined as the time it takes for plasma with thermal energy density $(3/2) P$ to lose this energy to a volumetric cooling power $Q_{\rm cool}^{-}$:
\begin{equation} \label{eq:tcool}
    t_{\rm cool} (z) \equiv \frac{3}{2} \frac{\langle P \rangle_{r \varphi t}}{\langle Q_{\rm cool}^{-} \rangle_{r \varphi t}}.
\end{equation}
Cooling, wind-outflow, and orbital timescales for the ZNF $\beta_0 = 10$ and NF $\beta_0 = 10^4$ simulations are compared in Figure~\ref{fig:timescales}. 

In the ZNF simulations, the cooling time is always shorter than the wind outflow time at $|z| = 3 H_z$. The leftmost two panels of Figure~\ref{fig:thermal_instability_ZNF} demonstrate how cooling dictates the $\rho$--$T$ distributions. As long as $t_{\rm cool} < t_{\rm wind}$, we see extended density distributions, crossing more than 0.5~dex in log-density space. However, extended temperature distributions only emerge when the orbital timescale is shorter than the cooling time. When $t_{\rm orb} < t_{\rm cool}$, broad temperature distributions develop, with temperatures from $10T_0$ all the way down to $T_0$. For $\mathcal{A} = 10^2$, where $t_{\rm cool} < t_{\rm orb}$ at $|z| = 3 H_z$, cooling is so rapid as to prevent the coronal temperature from rising much above $T_0$. However, once $t_{\rm orb} < t_{\rm cool}$, as is the case in the $\mathcal{A} = 10$ simulation, temperatures can rise to ${\sim}10T_0$ and gradually cool back down. As is clear from the upper-left and lower-centre panels of Figure~\ref{fig:thermal_instability_ZNF}, the $\rho$--$T$ distributions cross constant entropy surfaces while falling along lines of approximately constant pressure: a signature of isobaric cooling. Note that these panels correspond to situations where the plasma is thermal pressure dominated, i.e., $\langle \beta^{-1} \rangle_{r \varphi} < 1$ (Figure~\ref{fig:magnetization_alpha}), such that `isobaric' cooling implies approximately constant \textit{thermal} pressure rather than total, i.e. thermal plus magnetic, pressure.

When $t_{\rm orb} < t_{\rm wind} < t_{\rm cool}$, as is the case at $|z| = 5 H_z$ in the ZNF simulations, temperature remains confined around a narrow range from $(5$--$10)T_0$ as material does not have enough time to cool before being advected two scale heights upwards by the wind. For heights $|z| \geq 4 H_z$, material leaves the domain within a time $t_{\rm wind}$, resulting in a well-defined clustering of points (rightmost panels of Figure~\ref{fig:thermal_instability_ZNF}).

Perhaps the most revealing evidence of multiphase structure comes from the density-temperature distributions at $|z| = 4 H_z$ (centre column of Figure~\ref{fig:thermal_instability_ZNF}). Here, we see that, when $t_{\rm orb} < t_{\rm wind} < t_{\rm cool}$ as in the ZNF $\mathcal{A} = 10$ simulation, material clusters tightly around a narrow range of densities and temperatures, distributed along a constant pressure surface but not extending across a wide range of entropies. On the other hand, when $t_{\rm orb} < t_{\rm cool} < t_{\rm wind}$, cooling acts unabated, forming plasma with a wide range of temperatures from $T$~$=$~12~$T_0$ down to $T_0$. The densities of this cooling material extend across 0.7~dex in log-density space, all at nearly constant pressure. 

NF simulations provide insight into the regime of strong cooling, where $t_{\rm cool} < t_{\rm orb}$, and strong winds, where $t_{\rm wind} < t_{\rm orb}$. The strong-cooling regime is explored throughout Figure~\ref{fig:thermal_instability_NF}. Here, we see that the condition for strong cooling, $t_{\rm cool} < t_{\rm orb}$, is sufficient to allow the development of distributions broad in density but narrow in temperature, as exhibited by the $\mathcal{A} = 10^2$ NF simulations at heights $|z| = 3 H_z$, $4 H_z$, and $5 H_z$. At $5 H_z$, where $t_{\rm cool} \sim t_{\rm orb}$, a wider range of temperatures becomes accessible, yet temperatures still remain close to $T_0$ because cooling is still quite rapid. 

In the regime of strong wind driving, where $t_{\rm wind} < t_{\rm orb}$, the density-temperature distributions look dramatically different from even the broadened distributions in the ZNF simulations. At $|z| = 4 H_z$ and $5 H_z$ in the $\mathcal{A} = 10$ weak NF simulations, we find broad distributions covering a range of ${\sim}(1$--$10)T_0$ in temperature and similarly around 0.7~dex in log-density. However, unlike the ZNF distributions, distributions in the strong wind regime of the NF simulations do not fall along lines of nearly constant pressure; rather, they are skewed across lines of constant pressure \textit{and} constant entropy. 

These skewed NF distributions are not formed by cooling nor by adiabatic processes, and in fact, they would seem to imply that hotter material is on average more dense. One possibility for forming these skewed distributions is that energy is preferentially dissipated in dense material. However, another, perhaps more straightforward, means of achieving these skewed distributions is for the wind to advect hotter material rapidly from lower, denser regions of the corona up to the height at which distributions are measured. In this way, the temperature inversions in NF simulations (Figure~\ref{fig:temperature_density}) may be shaped by rapid outflows, which advect material heated lower in the corona to higher regions with lower mean density. Assessing whether advection or preferential heating in higher density material forms these skewed distributions requires a measurement of the heating rate, not cooling rate, as a function of density. 

\subsection{Impact of magnetic-field structure on conduction}
\label{sec:conduction}

While thermal conduction is \textit{not} included in our simulations, temperature inversions establish a situation in which conduction could transport heat from the hot corona into the colder disc. Here, we provide estimates for the heat flux from the corona into the disc expected for field-aligned conduction subject to our measured temperature gradients. The \cite{Spitzer1956} ion heat flux in the direction $\eb \equiv \boldsymbol{B}/|\boldsymbol{B}|$ in a corona with ion density $\rho$, pressure $P$, and temperature $T$ is
\begin{equation} \label{eq:Q_Spitzer}
    \boldsymbol{Q}_{\rm con}^{\rm Spitzer} = \frac{3.9 P}{L_T} \frac{k_{\rm B} T}{m_i} t_{ii} \eb,
\end{equation}
where $t_{ii}$ is the ion--ion collision time (see \S\ref{sec:thermo_timescales}) and we have introduced the thermal gradient length scale, $L_T \equiv -(\mathrm{d} \ln{T}/ \mathrm{d} z )^{-1}$. Similarly, the free-streaming heat flux as expected for saturated conduction is
\begin{equation} \label{eq:Q_free}
    \boldsymbol{Q}_{\rm con}^{\rm free} = \rho c_{s}^3 \: {\rm{sgn}} \left( (\ez \bcdot \eb) L_T \right) \eb. 
\end{equation}
We find that the Spitzer heat flux $\boldsymbol{Q}_{\rm con}^{\rm Spitzer}$ exceeds the free streaming heat flux $\boldsymbol{Q}_{\rm con}^{\rm free}$ when
\begin{equation} \label{eq:free_stream_condition}
    \mathcal{A} \lesssim 10 \: \frac{H_z}{L_T} \left( \frac{T_i}{T_e} \right)^{3/2} \left( \frac{m_e}{m_i}  \right)^{1/2} \left( \frac{n_{\rm i,c}}{n_0} \right)^{-1},
\end{equation}
where we have used the fact that the Coulomb coupling parameter $\mathcal{A}$ provides a measure of the ion--electron temperature equilibration timescale $t_{\rm eq}$. Note that for both heat fluxes, we are referring to \textit{ion} conduction, not electron conduction, the latter of which is subdominant since $T_i \gg T_e$ \citep{Spruit2002}. Using the estimate for $T_i/T_e$ in \S\ref{sec:thermo_timescales}, the approximate inequality \eqref{eq:free_stream_condition} becomes
\begin{equation} \label{eq:free_stream_A}
    \mathcal{A} \lesssim 10^3 \: \frac{H_z}{L_T} \left( \frac{n_{i,c}}{n_0} \right)^{-1}.
\end{equation}
In ZNF simulations, the minimum thermal gradient scales are typically around $0.5 \lesssim |L_T|/H_z \lesssim 0.8$, while NF simulations lie in the range $1 \lesssim |L_T|/H_z \lesssim 2$. Densities above $|z| = 2 H_z$ are below $\rho/\rho_0 \approx 0.1$ in all runs: the coronae of all simulations are in the free-streaming regime.

Both the free-streaming and Spitzer fluxes only apply along field lines. Across the field, the heat flux is suppressed by the ratio of the ion mean free path to gyroradius squared, which is enormous in strongly magnetized coronae. To determine how much the averaged vertical heat flux from the hot corona into the cold disc is decreased by the turbulent magnetic-field geometries that are realized in our simulations, we define a horizontally averaged suppression factor,
\begin{equation} \label{eq:f_suppression}
    f_{\rm s} \equiv \frac{\Big< \hat{\bb{z}} \bcdot \eb \left( \eb \bcdot \grad T \right) \Big>_{r \varphi t}}{\langle \hat{\bb{z}} \bcdot  \grad T \rangle_{r \varphi t}},
\end{equation}
which allows us to define a conduction timescale, 
\begin{equation} \label{eq:tcond}
    t_{\rm cond} \equiv \frac{3}{2} \frac{\langle P \rangle_{r \varphi t} |L_T|}{f_{\rm s} \langle \rho c_{\rm s}^3 \rangle_{r \varphi t}}= \frac{3}{2} \frac{\langle P \rangle_{r \varphi t} \langle T \rangle_{r \varphi t}}{\langle \rho c_{\rm s}^3 \rangle_{r \varphi t}} \left| \Big< \hat{\bb{z}} \bcdot \eb \left( \eb \bcdot \grad T \right) \Big>_{r \varphi t} \right|^{-1}.
\end{equation}
The conduction timescale is comparable to the orbital timescale, but prolonged by the inverse of the suppression factor, $f_{\rm s}^{-1}$,
\begin{equation}
    t_{\rm cond} \approx 0.2 f_{\rm s}^{-1} \frac{|L_T|}{H_z} \left( \frac{T}{T_0} \right)^{-1/2} \: \rm{orbits}.
\end{equation}

\begin{figure}
\hbox{
\includegraphics[width=0.48\textwidth]{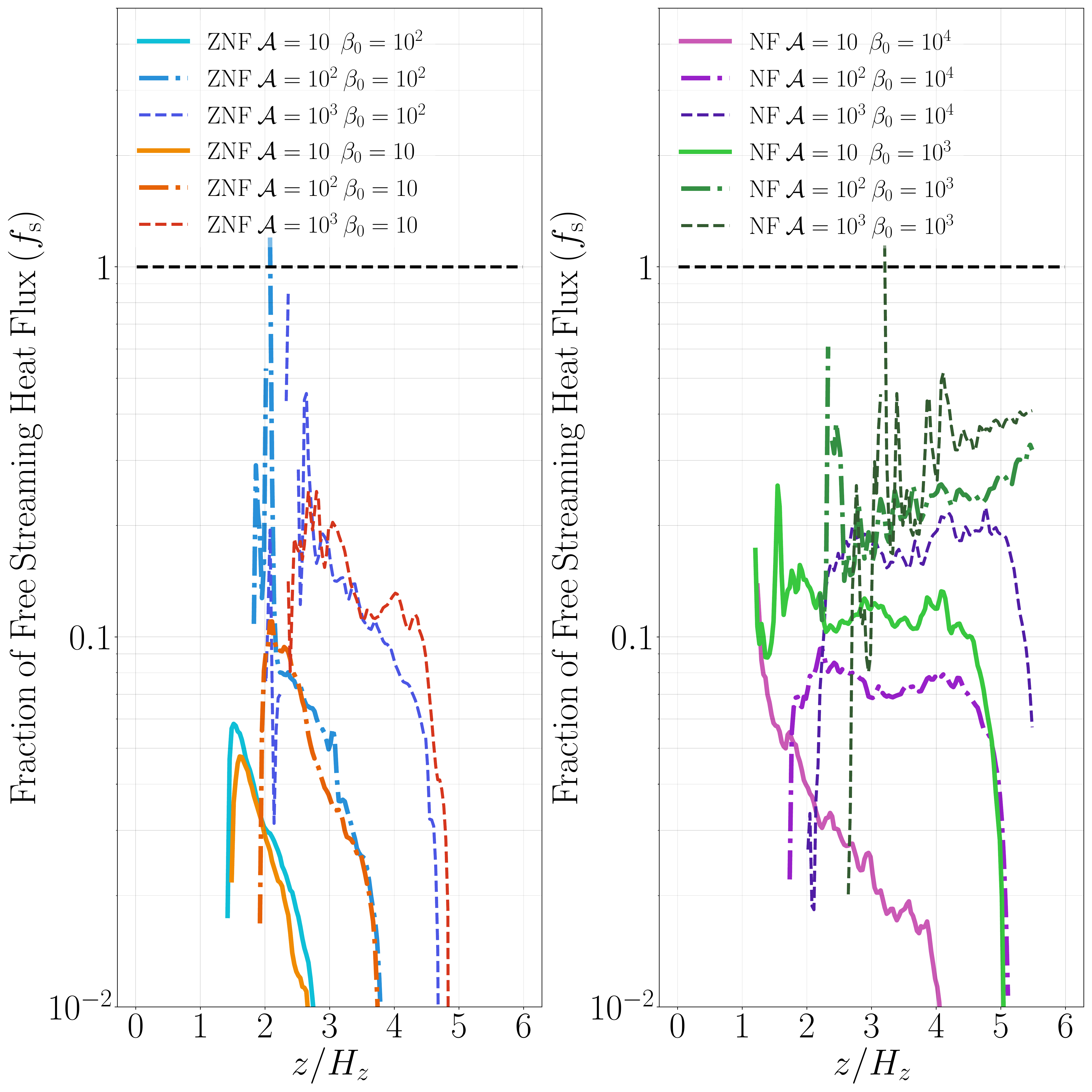}
}
\caption{Profiles of the conductive suppression factor $f_{\rm s}$ for all simulations, averaged from 30--100 orbits. The suppression factor represents a fraction of the free-streaming conductive heat flux ($f_{\rm s} = 1$ on this plot; black dashed line), and curves are only shown where the field-aligned temperature gradient and vertical temperature gradient are both in the same direction and positive. 
}
\label{fig:fs_profiles}
\end{figure}

Figure~\ref{fig:timescales} plots $t_{\rm cond}$ against the wind outflow and cooling timescales for all $\beta_0 = 10$ ZNF and $\beta_0 = 10^4$ NF simulations. The geometric suppression factor $f_{\rm s}$ is plotted in Figure~\ref{fig:fs_profiles}. In computing $f_{\rm s}$, we have first horizontally averaged the field-aligned vertical temperature gradient (the numerator of Equation~\ref{eq:f_suppression}) based on 700 output files from 30--100 orbits. The vertical temperature gradient (the denominator of Equation~\ref{eq:f_suppression}) is then computed by taking the $z$-derivative of horizontally averaged temperature profiles from each of these outputs and averaging these gradients in time. Figure~\ref{fig:fs_profiles} only shows regions where the temperature is increasing with increasing $z$, such that the heat flux would be into the disc, and where the numerator and denominator of Equation~\ref{eq:f_suppression} share the same sign. We remove measurements from $|z| > 5.5 H_z$, which are strongly influenced by the vertical boundaries. In addition, we omit contributions where $\rmd T/\rmd z < 10^{-2} (T_0/H_z)$, thus removing regions near the midplane where small, turbulent temperature variations around $T_0$ could generate some field-aligned temperature gradients. Without removing these small gradients in the denominator of Equation~\ref{eq:f_suppression}, $f_{\rm s}$ can blow up to be ${\gg}1$ near $|z| = H_z$, which is unphysical.

The heat flux from the corona into the disc would be substantially inhibited by the fields in these local simulations, particularly when $\mathcal{A} \lesssim 10^2$. While the conduction timescale $t_{\rm cond}$ is only a few times longer than the orbital timescale, two-temperature cooling is much more rapid at all heights in the corona. Figures~\ref{fig:timescales} and~\ref{fig:fs_profiles} display a few notable trends. (1) There is greater suppression of the field-aligned heat flux (i.e., $f_{\rm s}$ is smaller) with decreasing $\mathcal{A}$ in both ZNF and NF simulations, for fixed initial field. (2) Heat fluxes from the corona toward the disc extend deeper into the disc with decreasing $\mathcal{A}$, likely a consequence of the decrease of $z_T$ with decreasing $\mathcal{A}$. (3) NF simulations generally exhibit a wider range over which conduction can channel heat toward the disc, although only in the $\mathcal{A} = 10$ cases does the heat flux reach deep into the disc (near $|z| = H_z$). 

\section{Thermal effects on wind-driven accretion}
\label{sec:wind_accretion}

Accretion results from the outward transport of angular momentum. This transport can be provided by turbulence driven by the MRI and/or a wind that carries away angular momentum and exerts a torque on the disc. In this section, we study the loading and powering of accretion disc winds, focusing on the strong magnetocentrifugal winds launched by NF fields rather than the weak, thermally driven outflows in ZNF simulations (\S\ref{sec:thermally_driven_winds}). By scanning over box height for our NF $\mathcal{A} = 10$, $\beta_0 = 10^4$ run, we show that the mass-outflow rates from our weakly cooled, two-temperature models depend only weakly on box height. 

\subsection{Weaker cooling enhances mass and energy outflow rates}

\begin{figure}
\hbox{
\includegraphics[width=0.48\textwidth]{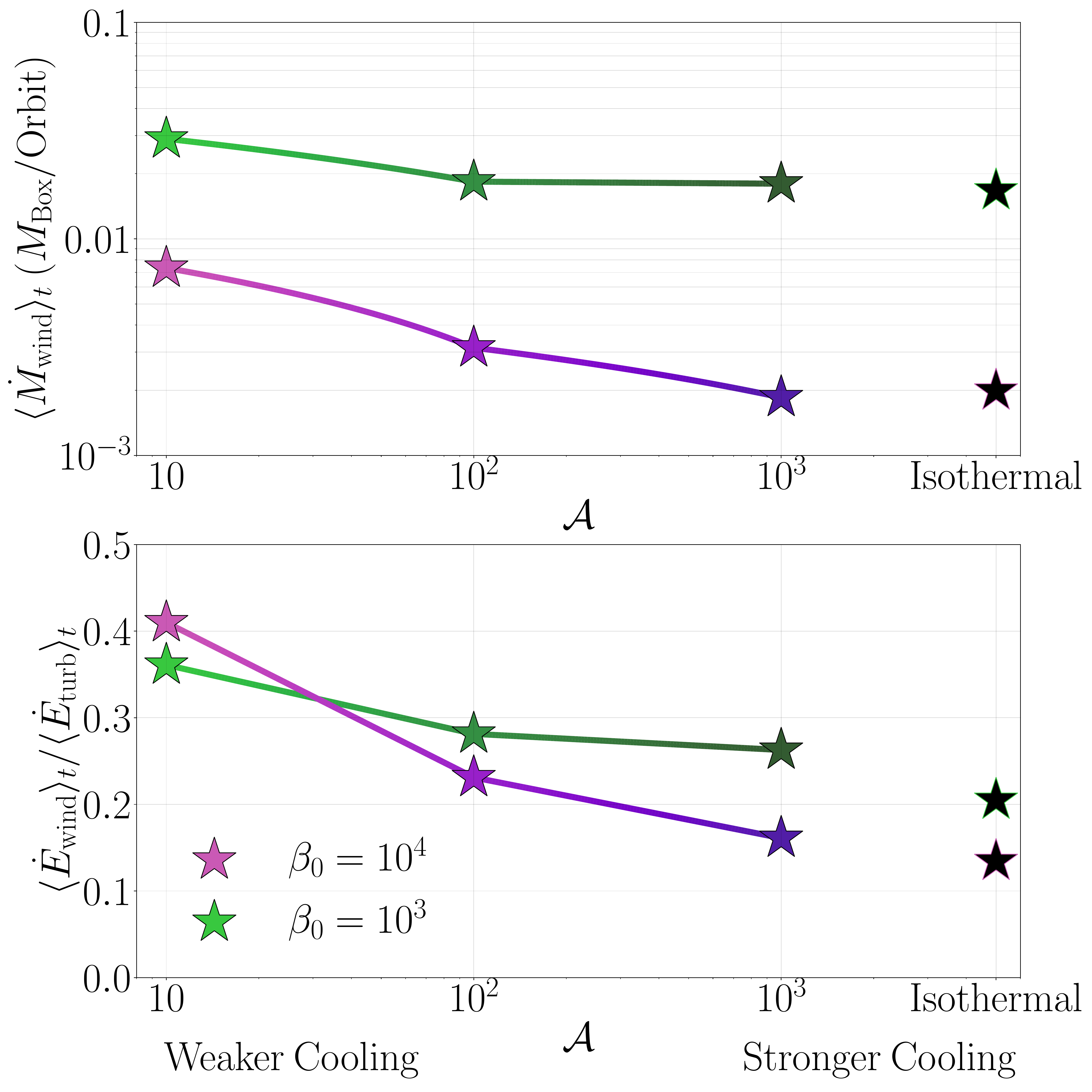}
}
\caption{Wind mass-outflow rates ($\dot{M}_{\rm wind}$) and energy-outflow rates ($\dot{E}_{\rm wind}$) as a function of cooling parameter $\mathcal{A}$ for all net flux simulations. Weaker cooling---smaller $\mathcal{A}$---enhances both mass and energy outflow rates. The effect of cooling is more pronounced on the weak field ($\beta_0 = 10^4$) simulations compared to the moderate field ($\beta_0 = 10^3$) runs. 
}
\label{fig:wind_mass_energy}
\end{figure}

Figure~\ref{fig:wind_mass_energy} displays the mass- and energy-outflow rates for all NF simulations. While mass is continually injected to maintain the initial box mass, we use the depletion time $t_{\rm dep}$ (Equation~\ref{eq:tdep}) as a proxy for the survival time of the global disc represented by our local simulations. Thermally driven ZNF winds are weak, and the underlying accretion disc can survive for more than 1500 orbits, $15$ times longer than our simulation run time. However, when a net-vertical flux threads the disc, winds can be quite strong, emptying the disc of material in anywhere from 570 orbits ($\beta_0 = 10^4$) to 70 orbits ($\beta_0 = 10^3$) in isothermal simulations, for the given parameters (box dimensions, mass-injection method, etc.) chosen for our study (Table~\ref{table:simulations}). 

Comparing the time-averaged mass-outflow rates $\langle \dot{M}_{\rm wind} \rangle_t$ between the $\mathcal{A} = 10$ and isothermal NF simulations, we find that weak cooling increases mass-outflow rates by as much as a factor of 3.8 when the field is weak ($\beta_0 = 10^4$), and a factor of 1.8 for the moderate-field ($\beta_0 = 10^3$) case. These are order-unity corrections. Mass-outflow rates depend more on the initial vertical magnetic flux threading the disc than on two-temperature thermodynamics.

Energy-outflow rates are similarly modified by an order-unity correction when the isothermality assumption is suspended. Again, comparing the $\mathcal{A} = 10$ and isothermal NF runs, we find that the wind outflow efficiency $\langle \dot{E}_{\rm wind} \rangle_t/\langle \dot{E}_{\rm turb} \rangle_t$ increases by a factor of ${\approx}2.9$ in the weak-field case and ${\approx}1.7$ for the moderate field. Yet, unlike mass-outflow rates, the efficiency with which energy is removed from the system depends more on thermodynamics than on the initial vertical magnetic flux. We caution though that stratified shearing boxes cannot always adequately constrain outflow rates because the efficiency of the wind can depend on the chosen box height. This point is  addressed further in the following sections.

\subsection{Critical points}
\label{sec:critical_points}

The mass outflow rate is determined at the slow magnetosonic point $z_{\rm s}$ \citep{Spruit1996}, which is defined as the point where the vertical velocity exceeds the slow magnetosonic speed $v_{\rm s}$ \citep{Ogilvie2012, Lesur2013}, 
\begin{equation}
    v_{\rm s}^2 = \frac{1}{2} \left[ v_{\rm A}^2 + c_{\rm s}^2 - \sqrt{ \left( v_{\rm A}^2 + c_{\rm s}^2 \right)^2 - 4 c_{\rm s}^2 v_{{\rm A},z}^2 } \right].
\end{equation}
Similarly, the energy outflow rate is determined at the fast magnetosonic point (hereafter, `fast point'), where the vertical velocity exceeds the fast magnetosonic speed $v_{\rm f}$,
\begin{equation}
    v_{\rm f}^2 = \frac{1}{2} \left[ v_{\rm A}^2 + c_{\rm s}^2 + \sqrt{ \left( v_{\rm A}^2 + c_{\rm s}^2 \right)^2 - 4 c_{\rm s}^2 v_{{\rm A},z}^2 } \right].
\end{equation}
Above, $v_{\rm A} \equiv |\boldsymbol{B}|/\sqrt{\rho}$ is the \alf speed, $c_{\rm s} \equiv \sqrt{\gamma P/\rho} = \sqrt{\gamma T}$ is the adiabatic sound speed, and $v_{{\rm A},z} \equiv B_z/\sqrt{\rho}$ is the \alf speed along the mean $z$-field. In practice, we take the slow point as the location $z_{\rm s}$ where $\langle v_z \rangle_{r \varphi t} (z_{\rm s}) = \langle v_{\rm s} \rangle_{r \varphi t} (z_{\rm s})$ and similarly, for the fast point $z_{\rm f}$, $\langle v_z \rangle_{r \varphi t} (z_{\rm f}) = \langle v_{\rm f} \rangle_{r \varphi t} (z_{\rm f})$. The slow points are indicated by closed circles in Figures~\ref{fig:temperature_density} and~\ref{fig:vz_profiles}.

Winds remove angular momentum by exerting a torque on the disc. For the range of NF magnetic field strengths explored in this work, \cite{Bai2013_NF} confirmed the magnetocentrifugal behaviour of winds in the shearing box. Along a given poloidal field line, the magnetic field exerts a torque on the wind material and therefore imparts angular momentum to the outflow. This tell-tale sign of a ``magnetocentrifugal'' wind persists despite the fact that the toroidal field dominates over that in the poloidal direction, and velocity streamlines are misaligned with the field lines \citep{Lesur2013}. 

The length of the lever arm that exerts this torque is set by the \alf point in the flow, i.e., the point where the wind velocity exceeds that of \alf waves propagating along poloidal field lines. In a lightly loaded wind with well-defined poloidal fields, this definition is apparent. However, winds in shearing boxes are turbulent \citep{Fromang2013}, and the relevant \alf speed should be determined by averaging over turbulent fluctuations at fixed height.  It was not a priori obvious to us that the turbulent transport in the vertical direction is negligible compared to the transport by the mean magnetic fields, particularly since the mean $B_z$ is a constant in the shearing box (and relatively small in our models).   However, explicitly calculating the turbulent contributions to the vertical transport shows that they are indeed small compared to the transport by the mean magnetic fields.  This is primarily because the mean toroidal field is very strong.\footnote{One subtlety is that this comparison  should only be done over a time interval with a fixed sign of $B_\phi$, i.e., between toroidal field reversals. Averaging over the latter would artificially suppress the mean $B_\phi$ contributing to angular momentum transport in the wind.}

Thus, we proceed by defining the \alf point in the traditional way as the point where the horizontally averaged vertical velocity reaches the \alf speed, $\langle v_z \rangle_{r \varphi t} (z_{\rm A}) = \langle v_{{\rm A},z} \rangle_{r \varphi t} (z_{\rm A})$. The horizontal average of the \alf point is defined as
\begin{equation} \label{eq:vAz_rphi}
    \langle v_{{\rm A},z} \rangle_{r \varphi t} = \frac{\langle B_z \rangle_{r \varphi t}}{\sqrt{\langle \rho \rangle_{r \varphi t}}},
\end{equation}
where, as was done with $\beta^{-1}$ (Equation~\ref{eq:beta_inv}), we have averaged the field and density separately to reduce the effects of isolated cells with enormous \alf speeds.

Slow points are always on the domain in our simulations, independent of thermodynamics and initial field configuration. At least for the ZNF $\mathcal{A} = 10$ and NF simulations, which attain sonic transitions away from the domain boundary, the measured mass-outflow rates are reliable. However, because the fast point is never on the domain in any of our simulations, we cannot constrain the energy-outflow rates using our local shearing-box simulations.

For NF simulations, the \alf point $z_{\rm A}$ is always located within the domain, between $2 \lesssim z_{\rm A}/H_z \lesssim 4.5$. This height increases modestly with increased cooling in the moderate-field runs (from $3.7 H_z$ to $4.3 H_z$), and by as much as a factor of 2 between $\mathcal{A} = 10$ and isothermal runs (from $2.2 H_z$ to $4.4 H_z$) for $\beta_0 = 10^4$. 

Appendix~\ref{sec:wind_survival} provides a comparison of the stress due to the wind ($\alpha_z \equiv \langle \mathcal{T}_{z \varphi}/P_0 \rangle$), which transports angular momentum in the vertical $z$ direction, with the turbulent $\alpha_{\rm mid}$ responsible for the radial transport of angular momentum. In principle, if $\alpha_z/2 \alpha_{\rm mid} > H_z/R_0$ for an assumed disc thickness $H_z/R_0$, then winds dominate angular momentum transport \citep{Fromang2013}. In practice, however, making this comparison requires an arbitrary choice of $|z|$ at which to compare $\alpha_z$ and $\alpha_{\rm mid}$. We find that $\alpha_z$ varies with height $|z|$ as some of the wind stress goes toward driving radial flows that approach the sound speed at high $|z|$. The presence of these radial flows indicates an exchange of angular momentum between the field and disc, consistent with a \cite{Blandford1982} magnetocentrifugal wind. Choosing, somewhat arbitrarily, to evaluate the wind stress at the \alf point, we find that, for discs thinner than $H_z/R_0 \sim 0.02$, the wind stress will dominate over the turbulent radial stress while, for thicker discs, the latter dominates the angular momentum transport.

\begin{figure}
\hbox{
\includegraphics[width=0.48\textwidth]{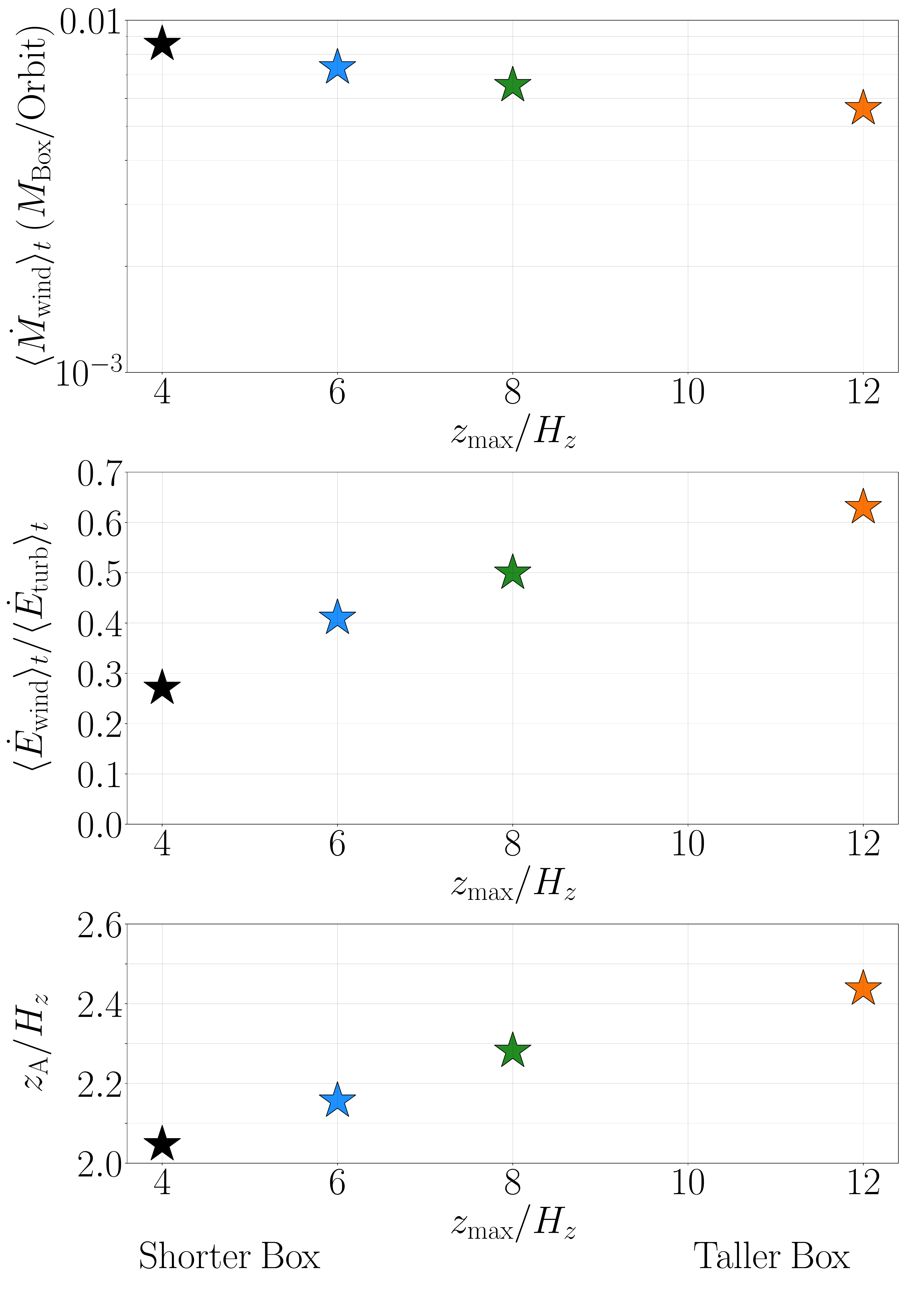}
}
\caption{Mass-outflow rate ($\dot{M}_{\rm wind}$; top), energy-outflow rate ($\dot{E}_{\rm wind}$; middle), and \alf point ($z_{\rm A}$; bottom) as functions of box height $z_{\rm max}$ (measured from the disc midplane) for NF simulations with $\mathcal{A} = 10$ and $\beta_0 = 10^4$. The same resolution (32 cells/$H_z$) is used for all simulations, and the time averages are all from $30$--$100$~orbits. Mass-outflow rates are nearly converged with $z_{\rm max}$ and the \alf point increases by ${\lesssim}H_z$ when tripling the box height. 
}
\label{fig:height_scan}
\end{figure}

\subsection{Convergence with box height}
\label{sec:height_convergence}

Mass-outflow rates from NF, isothermal, stratified shearing boxes are known to depend on the box height $z_{\rm max}$. By varying $z_{\rm max}$ for the weak NF ($\beta_0 = 10^4$) simulation with weak two-temperature cooling ($\mathcal{A} = 10$), we investigate if the mass- and energy-outflow rates in our two-temperature models have a similar dependence on box height. Our results are shown in Figure~\ref{fig:height_scan}.  

\begin{figure*}
\hbox{
\includegraphics[width=1.0\textwidth]{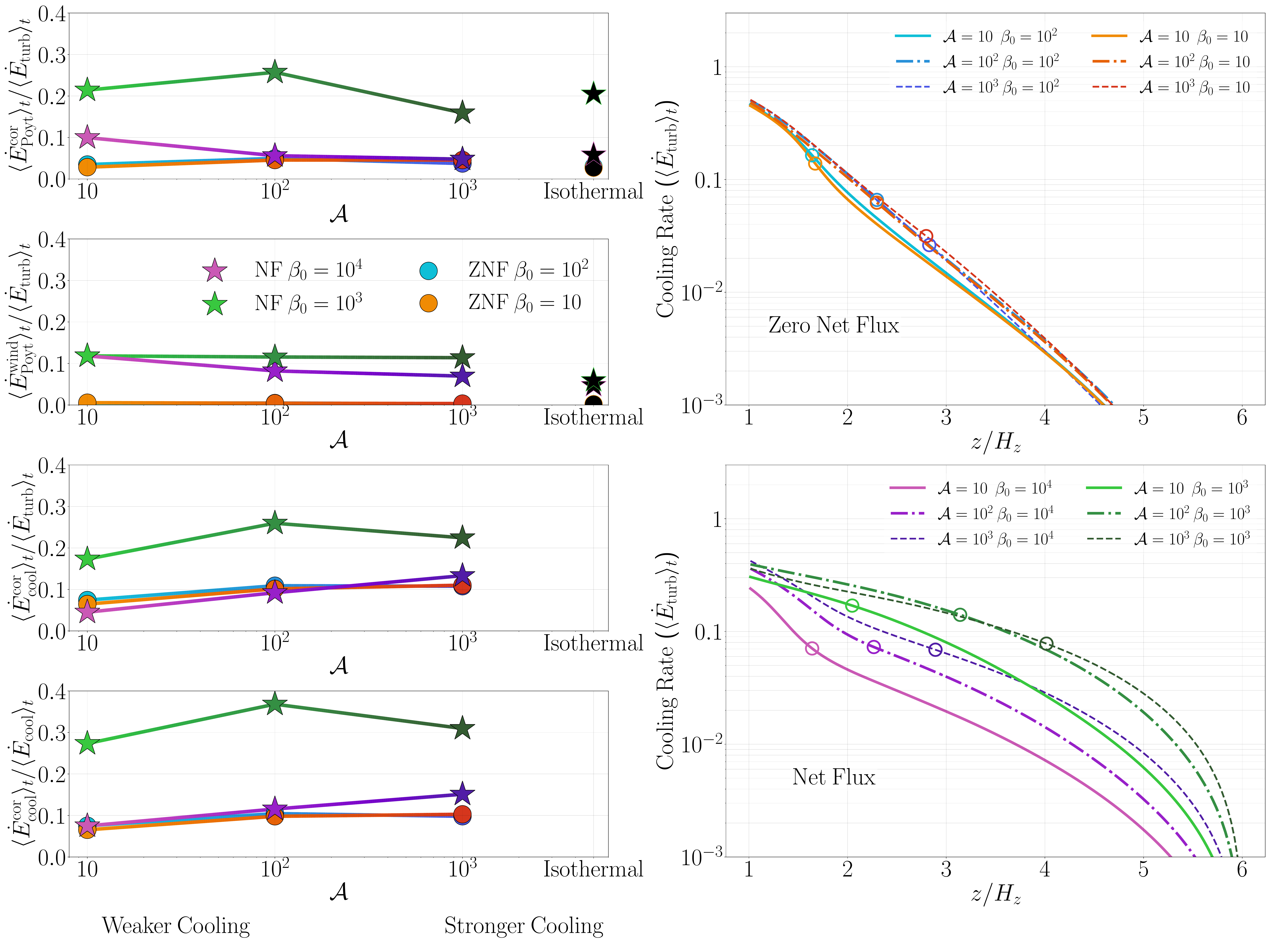}
}
\caption{From top to bottom left: Poynting flux into the corona $\dot{E}_{\rm Poyt}^{\rm cor}$ through $|z| = 2 H_z$, Poynting flux in the wind (Equation~\ref{eq:Poynting_wind}), cooling rate in the corona $\dot{E}_{\rm cool}^{\rm cor}$ (i.e. from $2 < |z|/H_z < 6$) relative to the power injected by the shearing boundary ($\dot{E}_{\rm turb}$), and cooling rate in the corona $\dot{E}_{\rm cool}^{\rm cor}$ relative to the box-averaged cooling rate, $\dot{E}_{\rm cool}$. All powers are averaged from 30--100 orbits. Right: Cooling efficiency (see text) as a function of height for all zero net-flux (ZNF; top right) and net-flux (NF; bottom right) simulations. Open circles correspond to cooling rates at $|z| = z_T$ (Table~\ref{table:simulations}). ZNF simulations generate weak Poynting fluxes that never exceed $\approx 5$ per cent of $\dot{E}_{\rm turb}$. However, moderate NF ($\beta_0 = 10^3$) runs generate far more powerful Poynting fluxes that thermalize above $|z| = 2 H_z$, releasing anywhere from $15-30$ per cent of the injected energy. While the cooling rate drops precipitously above $|z| = H_z$ in ZNF runs, NF simulations exhibit much flatter cooling efficiency profiles, implying that substantial energy reaches higher $|z|$ and is released there compared to ZNF simulations. 
}
\label{fig:cooling_efficiency}
\end{figure*}

Weakly cooled ($\mathcal{A} = 10$) two-temperature simulations show much better convergence with height $z_{\rm max}$ compared to isothermal simulations. \cite{Fromang2013} found that for a NF $\beta_0 = 10^4$ field, extending their simulation domain from a height $z_{\rm max} = 3.5 H_z$ to $7 H_z$ caused the mass outflow rate to decrease by a factor of $1.1 \times 10^{-3}/3.2 \times 10^{-4} \approx 3.4$ (comparing simulations \texttt{Tall4H} and \texttt{Diffu4H} from their table~1; note the factor of $\sqrt{2}$ difference in their definition of $H_z$, hence $20 H_{\rm Fromang} \approx 14 H_z$). Following the scaling found from their doubling of $z_{\rm max}$, the outflow rate between $z_{\rm max} = 4 H_z$ and $12 H_z$ should drop by a factor of $\approx 6$ if $\mathcal{A} = 10$ and isothermal runs behave similarly. Instead, $\dot{M}_{\rm wind}$ in our weak NF $\mathcal{A} = 10$ model drops by a more modest factor of ${\approx}1.5$ over the same tripling of the box height. Weak cooling improves the convergence properties of our simulations relative to isothermal runs, and thus, depletion times in our local two-temperature models may be more indicative of the behaviour of real, global systems.  If this is indeed the case, sufficiently thin discs are likely to evaporate to form radiatively inefficient accretion flows via the winds into their two-temperature coronae.  In Appendix~\ref{sec:wind_survival}, we estimate that this evaporation occurs for $H_z/R_0 \lesssim 0.1$ if the outflow rates found here are indicative of those present in global simulations.   Note though that convergence likely worsens with increasing $\mathcal{A}$ as the plasma behaves more and more like an isothermal fluid. Thus, the improved convergence exhibited by our NF $\mathcal{A} = 10$, $\beta_0 = 10^4$ simulation may not be a quality shared by runs with different $\mathcal{A}$ and field configurations.

While mass-outflow rates may converge well with box height, the same cannot be said of the wind efficiency $\dot{E}_{\rm wind}/\dot{E}_{\rm turb}$. Indeed, Figure~\ref{fig:height_scan} shows that $\dot{E}_{\rm wind}/\dot{E}_{\rm turb}$ can increase by as much as a factor of~2 when tripling the box height. Subsequently, for the box-integrated relation $\dot{E}_{\rm turb} = \dot{E}_{\rm wind} + \dot{E}_{\rm cool}$ to hold, increasing the efficiency of wind driving implies \textit{decreasing} the efficiency of cooling measured over the entire box. Note, though, that the efficiency of cooling in the corona alone is relatively insensitive to $z_{\rm max}$ (see Appendix~\ref{sec:height_scan}).

Energy fluxes in these winds are dominated by the gravitational energy flux, $\dot{E}_{\rm Grv} \equiv \oiint ( \Phi + \frac{3}{2} \Omega^2 x^2 ) \rho u_z \: \rmd x \rmd y$, in all but the shortest box with $z_{\rm max} = 4 H_z$, where the gravitational energy and Poynting fluxes are comparable. Of the ${\approx}64$ per cent of $\langle \dot{E}_{\rm turb} \rangle_t$ carried by the wind in the tallest box run, ${\approx}32$ per cent is in the gravitational energy flux and ${\approx}16$ per cent is in Poynting flux, with the remaining 16 per cent divided between thermal energy flux (10 per cent) and kinetic energy flux (6 per cent). Because the fast point is never on the domain, all wind energy fluxes increase with $z_{\rm max}$ (Appendix~\ref{sec:height_scan}). Interestingly, profiles of the vertical energy flux normalized by the turbulent injection rate $\langle \dot{E}_{\rm turb} \rangle_t$ are nearly identical above $|z| = 2 H_z$, independent of box height (at least between 30 and 60 orbits, before an asymmetry arises in the tallest box run, likely due to the stochastic behaviour of the toroidal field dynamo). This similarity in vertical energy flux profiles indicates that the box height simply sets the measurement point of the energy-outflow rate. 

\section{Poynting flux and cooling}
\label{sec:Poynting_cooling}

The left panels of Figure~\ref{fig:cooling_efficiency} display the time-averaged Poynting flux into the corona ($\langle \dot{E}_{\rm Poyt}^{\rm cor} \rangle_t$), the Poynting flux contribution to the wind ($\langle \dot{E}_{\rm Poyt}^{\rm wind} \rangle_t$), where
\begin{equation} \label{eq:Poynting_wind}
    \dot{E}_{\rm Poyt}^{\rm wind} = \oiint \boldsymbol{\mathcal{S}} \bcdot \hat{\bb{z}} \left( |z| = z_{\rm max} \right) \: \rmd x \rmd y,
\end{equation}
and the time-averaged rate of cooling in the corona above $|z| = 2 H_z$, i.e. $\langle \dot{E}_{\rm cool}^{\rm cor} \rangle_t$, for all simulations. We normalize these powers by the turbulent injection power, $\langle \dot{E}_{\rm turb} \rangle_t$ (Equation~\ref{eq:turbulent_injection}). 

In the right panels of Figure~\ref{fig:cooling_efficiency}, we show the cooling rate integrated above a height $z$. We horizontally and time average measurements of $Q^{-}_{\rm cool} (x,y,z)$ taken from 700 output data files between 30 and 100 orbits. The $+z$ and $-z$ profiles of $\langle Q^{-}_{\rm cool} \rangle_{r \varphi t}$ are then averaged together, and we integrate from $z$ to $z_{\rm max}$ to determine the total cooling power released above height $z$. 

The main result of our analysis is that ZNF simulations are inefficient at launching Poynting fluxes into the corona, with ${\lesssim}5$ per cent of the energy injected into the disc reaching the corona as a net Poynting flux. On the other hand, NF simulations are remarkably efficient at generating vertical Poynting fluxes, transporting anywhere from 10~per cent (in the case of the $\mathcal{A} = 10$, $\beta_0 = 10^4$ weak NF simulation) to 26~per cent (for the $\mathcal{A} = 10^2$, $\beta_0 = 10^3$ moderate NF run) of $\langle \dot{E}_{\rm turb} \rangle_t$ into the corona. 

Simulations with NF fields radiate a significant fraction of the accretion flow's power at large $z$, with moderate NF field simulations liberating up to 26 per cent of $\langle \dot{E}_{\rm turb} \rangle_t$ above $|z| = 2 H_z$, and weak NF field simulations reaching coronal cooling powers of 13~per cent of $\langle \dot{E}_{\rm turb} \rangle_t$ (\citealt{Scepi2023} found qualitatively similar results in global simulations of thin, magnetically arrested discs).  ZNF field runs are less efficient at radiating their energy via the corona compared to moderate NF field simulations, releasing about 11~per cent of the injected energy in the corona, almost independent of thermodynamics and initial field strength. The cooling rate in the coronae of ZNF simulations has only a weak dependence on thermodynamics, with a slight drop of 3--4 per cent of $\langle \dot{E}_{\rm turb} \rangle_t$ in the $\mathcal{A} = 10$ runs around $z = 2 H_z$ compared to more strongly cooled cases. However, NF simulations display a greater dependence on thermodynamics, with stronger Coulomb coupling (larger $\mathcal{A}$) implying higher cooling efficiencies at a given height for both weak and moderate NF fields. Notably, like the mass-outflow rates, the cooling efficiency depends more on the initial net vertical magnetic flux than the value of $\mathcal{A}$.

Crucially, NF simulations maintain higher cooling efficiencies at large $z/H_z$, with strongly cooled ($\mathcal{A} \gtrsim 10^2$) moderate NF simulations maintaining efficiencies above 10 per cent up to a height of $|z| = 4 H_z$. Similarly, weak NF simulations remain above efficiencies of 1~per cent up to $|z| = (3.5-5) H_z$. NF simulations remain efficient at releasing the underlying injected energy via cooling, even at large heights in the domain, while efficiencies in  ZNF simulations drop precipitously, reaching 0.3~per cent by $|z| = 4 H_z$. Because our simulations lack radiation transport, the height of the $\tau_{\rm es} = 1$ surface is arbitrary, and may be farther away from the midplane than $|z| = 2 H_z$. Thus, the more accurate definition of the `corona' in our simulations may be the region above $|z| = z_T$ (shown by open circles in Figure~\ref{fig:cooling_efficiency}). However, for the ZNF $\mathcal{A} = 10$ simulation, $z_T < 2 H_z$ such that the field loops are initially extending above $z_T$. A ZNF field confined to lower $|z|$ closer to the midplane may exhibit even weaker cooling in the corona (see discussion in \S\ref{sec:previous_work}).

We note that the cooling rates reported here do not include any direct electron heating physics. The right panels of Figure~\ref{fig:cooling_efficiency} simply show the energy that would be imparted to electrons via Coulomb collisions. Physical processes that lie beyond the purview of MHD, e.g., collisionless magnetic reconnection or turbulent dissipation, may directly heat electrons beyond what we infer from our modeling of Coulomb collisions;
%or increase the rate of dissipation in the corona. 
we return to this point in \S\ref{sec:efficient_heating}.

\section{Discussion}
\label{sec:discussion}

\subsection{Corona ion temperatures}
\label{sec:hot_corona}

Taking the coronal temperature to be a fixed fraction of $T_{\rm virial}$ allows us to express the ion temperature in the corona in a way that is independent of the midplane temperature $T_0$. This choice is justified because, at least for the NF $\mathcal{A} = 10$, $\beta_0 = 10^4$ simulation (see Appendix~\ref{sec:height_scan}), the peak horizontally averaged temperature is approximately ${\propto} z_{\rm max}^2$. Thus, the ratio of the peak temperature to the virial temperature $T_{\rm virial}$ (Equation~\ref{eq:T_virial}) is approximately independent of box height (and depends primarily on the strength of cooling, set by the Coulomb coupling parameter $\mathcal{A}$, and the magnetic-field configuration). 

Using the peak temperature $\langle T \rangle_{r \varphi t} \approx 4 T_0$ attained in the moderate NF ($\beta_0 = 10^3$) $\mathcal{A} = 10$ simulation and the virial temperature for the fiducial box height, $T_{\rm virial} = 24 \: T_0$ (Equation~\ref{eq:T_virial}), we find the temperature of the coronal ions to be $T_{\rm ion} {\approx } (1/6) T_{\rm virial}$. Based on the estimate from \S\ref{sec:thermo_timescales}, this temperature corresponds to $T_{\rm ion} \approx 6 \times 10^{10} \: \rm{K}$! While radiation is necessary to accurately assess the disc's temperature and density structure, by focusing our attention on the ion thermodynamics alone, our two-temperature model is able to form a hot corona with temperatures well in excess of $10^9$ K. 

\subsection{Comparisons to previous work}
\label{sec:previous_work}

Temperature inversions in stratified shearing-box simulations are not inevitable; rather, they are a consequence of our assumed physics. Here, we discuss two alternative approaches to treating disc thermodynamics -- radiation MHD and a variable adiabatic index $\gamma$ -- with a focus on how our conclusions compare to previous work.

Early radiation-MHD simulations by \cite{Hirose2006} did not form temperature inversions. Instead, radiation diffusion through their optically thick disc ``puffed up'' the disc and lifted dense material into the corona, inhibiting the rise of the coronal temperature. Similarly, \cite{Jiang2014} found that temperature inversions are a consequence of a lower midplane density, which allows radiation to escape without puffing up the disc. 

Both \cite{Hirose2006} and \cite{Jiang2014} focused on ZNF fields. In the case of \cite{Jiang2014}'s simulation~A, which did form temperature inversions, the ratio of the horizontally averaged coronal temperature to the midplane temperature is similar to that obtained in our ZNF simulations (Figure~\ref{fig:temperature_density}). Notably though, their corona, defined self-consistently as the region above the surface where the optical depth to electron scattering $\tau_{\rm es}$ crosses 1, is near $|z| = H_z$ (see their figure~1, top panel). This is closer to the midplane than our assumed (and arbitrary) disc-corona transition of $|z| = 2 H_z$. Interestingly, the dissipation fraction in their corona, which they estimate to be about 3.4~per cent of the injected energy, is smaller than what we estimate the cooling rate to be above $|z| = 1$: 40--50 per cent, based on the cooling rate profiles (Figure~\ref{fig:cooling_efficiency}, right panels). The key difference between our simulations and that of \cite{Jiang2014} is that the magnetic fields of \cite{Jiang2014} are initially confined to $|z| < 0.8 H_z$, not $|z| < 2 H_z$, as in our calculations. Thus, consistent with our findings in \S\ref{sec:Poynting_cooling}, ZNF fields are inefficient at transporting energy above the heights within which they are initially confined, at least in local simulations. If this behaviour of ZNF fields applies in global systems, then the primary means for releasing high fractions of the flow's power via the optically thin corona is to decrease the height of the $\tau_{\rm es} = 1$ surface such that dissipation is concentrated in optically thin plasma. 

NF fields generate substantially larger Poynting fluxes into the corona than ZNF fields, offering an alternative mechanism for efficient coronal heating. The effects of thermodynamics on local accretion disc simulations with NF fields were first explored by \cite{Io2014}, who examined a much weaker NF field ($\beta_0 = 10^6$) using a taller box ($z_{\rm max}= 16 H_z$), lower resolution (16~cells/$H_z$ rather than our 32~cells/$H_z$), and treated thermodynamic uncertainties using a variable adiabatic index $\gamma$ in the range $1 \leq \gamma \leq 5/3$, rather than a cooling function. These authors found temperature inversions in all of their models other than the isothermal ($\gamma = 1$) reference run, with higher peak temperatures attained for higher $\gamma$. Because their NF fields were so weak, thermal driving dominated over magnetocentrifugal acceleration, the latter being the primary mechanism powering winds in our NF simulations. 

\subsection{Multiphase structure}
\label{sec:multiphase_discussion}

When $t_{\rm orb} < t_{\rm cool} < t_{\rm wind}$ in the ZNF simulations, we find evidence for multiphase structure, with broadened density and temperature distributions forming along curves of approximately constant thermal pressure (Figure~\ref{fig:thermal_instability_ZNF}). In real systems, we expect that the strong density dependence ($\propto \rho^2$) of Coulomb collision-mediated cooling will act to form rich, multiphase structure, with colder clumps embedded in a hot, virialized, corona.

Similar structures have been observed in simulations of thermally driven winds launched from radii much further away from the black hole than $10 R_g$. \cite{Dannen2020} found clumpy structures could form in global axisymmetric hydrodynamic simulations of parsec-scale outflows, so long as the gas entered a thermal instability zone and rapid acceleration of the wind did not negate thermal instability through the stretching of unstable entropy modes \citep{Waters2021}. Similarly, \cite{Waters2022} showed that multiphase gas can form at larger radii in the outflows as gas passes through a thermal instability zone and circulates in large scale vortices produced by the flow. However, at smaller radii, rapid acceleration of the wind tends to negate thermal instability. 

While we have proposed criteria for the formation of multiphase gas using similar techniques as those applied to dynamical thermal instability \citep{Balbus1986, Waters2021}, the multiphase structure viewed in our simulations is quite distinct from that observed in these hydrodynamic simulations. First and foremost, our simulations are focused on radii very close to the ISCO of the black hole. Thermally driven winds are thought to be driven by Compton heating of discs at larger radii \citep{Begelman1983}, where the source of the Comptonizing photons is the corona investigated in our work. Further, the strong MRI-driven turbulence responsible for facilitating transport through our model coronae can produce density and temperature structure that is difficult, if not impossible, to disentangle from structure produced by thermal instability. Our weak NF $\mathcal{A} = 10^2$ simulations (Figure~\ref{fig:thermal_instability_NF}) do not show a broad range of temperatures because $t_{\rm cool} < t_{\rm orb}$ at all radii except $|z|/H_z \gtrsim 5$, so the broad density distributions may be as much a result of MHD turbulent fluctuations as two-temperature cooling. When $t_{\rm wind} < t_{\rm orb}$, density-temperature distributions in our simulations are skewed toward having more hot material at higher densities, which may be an effect of how we are measuring these distributions at fixed $|z|$, rather than a consequence of cooling, which is slow, particularly for the weak NF $\mathcal{A} = 10$ simulations at heights $|z|/H_z \gtrsim 4$. Finally, the thermally driven winds produced by the ZNF simulations are quite distinct from Compton heated winds as the heating driving the winds is from the dissipation of MHD turbulence, and the winds are much weaker than large scale thermally driven winds, barely attaining a sonic transition on our domain (Figure~\ref{fig:vz_profiles}).

This picture becomes even more complex when we consider the fate of leptons. If leptons are left to cool at the Compton rate, uninhibited by direct or Coulomb collisional heating, we should expect that coronae will have regions of much lower electron temperature and much higher electron density, surrounded by hot, diffuse, ion-pressure-supported regions. In these more dense regions, $\Omega t_{\rm eq} \ll 1$ and ions will begin to cool at the Compton rate as the plasma behaves like a one-temperature fluid. Thus, the widths of the density and temperature distributions found in our simulations may be underestimates of the true multiphase nature of realistic coronae. The spatial distribution of dense vs. rarified regions may depend sensitively on the spatio-temporal intermittency of ion vs. electron heating, which is shaped by kinetic processes. 

Multiphase structure has a number of important observable consequences. In one sense, the continual formation and destruction of optically thick and thin regions in the hot coronal atmosphere is likely to influence the variability of the corona, as observed in hard X-rays. Similarly, if the corona is clumpy or patchy, the covering fraction of the corona relative to the underlying disc may be modified. Most notably, decreasing the covering fraction of the corona may allow soft flux produced by disc reflection \citep{George1991} to escape the system without being upscattered in the corona \citep{Wilkins2015}, which in a standard ``sandwich'' corona, might otherwise lead to a softening of the photon index to the point of discord with observations \citep{Dove1997_sandwich}. Finally, runaway ion cooling may allow for condensation out of the corona which can provide an optically thick iron line signal at the ISCO, even if the disc is truncated \citep{Liska2022}. Work which can accurately capture the interplay of ion--electron temperature equilibration, Compton cooling, pair production, and differential electron and ion heating imposed by kinetic processes is necessary to resolve these questions.  

\subsection{Wind-driven accretion and disc evaporation}

Magnetocentrifugal winds, such as those launched in our NF simulations (\S\ref{sec:wind_accretion}), exert a magnetic torque on the disc and enhance the rate of accretion. Simultaneously, the winds act to deplete the disc of material, emptying the entire disc of plasma on a timescale $\sim t_{\rm dep}$.  The latter is of order hundreds of orbits in our simulations (Table \ref{table:simulations}).

In a global system where self-consistent radial inflow is necessary to offset disc winds, accretion, driven both by winds and by internal stresses, competes with depletion to set the disc density structure. At present, it is unclear which process---radial accretion vs. wind depletion---wins the competition. Likely, for thicker discs with larger $H_z/R_0$, accretion is more rapid than wind depletion: the viscous time is $t_{\rm visc} \sim (R_0/H_z)^2 \left( \alpha_{\rm mid} \Omega \right)^{-1}$. However, for thinner discs, wind-driven depletion may act to reduce the disc density, particularly in the surface layers from which the wind originates. This reduced density increases $\Omega t_{\rm eq}$, potentially to the point where ion and electron temperatures decouple, and runaway ion heating evaporates the thin disc. This process may further be complicated by non-axisymmetric effects inaccessible to our local models, where one section of the disc (in azimuth, $\varphi$) evaporates before other sections follow suit \citep{Proga2005}. In Appendix~\ref{sec:wind_survival}, we provide a rough estimate for the transition between these two fates  using our shearing box outflow rates as a guide to the outflows that might be present in global systems. Our local simulations suggest that global discs with $H_z/R_0 \lesssim 0.1$ are prone to evaporation into a radiatively inefficient accretion flow (RIAF) because the outflow rate into the corona is faster than the inflow to smaller radii. Whether or not this process would indeed proceed unabated, leaving behind a RIAF in place of a thin disc, is left as a subject for future investigations. 

\subsection{Efficient coronal heating by net flux fields}
\label{sec:efficient_heating}

In our simulations, NF fields are far more efficient at generating net Poynting fluxes into the corona compared to ZNF fields (\S\ref{sec:Poynting_cooling}). How this energy is eventually dissipated though, remains an open question. 

Resolving the mechanisms by which ZNF and NF fields heat coronae provides motivation for future studies on the connections between mesoscale MHD simulations and miroscale kinetic processes. For instance, in the two-temperature solar corona, closed magnetic field loops, i.e. ZNF fields, release their energy primarily through small-scale, collisionless magnetic reconnection \citep{Drake2006, Zweibel2009, Sironi2014, Guo2014}, which powers flaring events. On the other hand, in models for heating the open-field solar corona, convective motions at the solar surface launch Alfv$\acute{\text{e}}$nic wave packets out along field lines, and some fraction of the wave energy is reflected back inward, toward the solar surface. The nonlinear interactions between inward- and outward-traveling waves initiates a turbulent cascade, which heats the corona \citep{Heinemann1980, Dmitruk2002, Chandran2009}. This mechanism has gained recent attention in the context of accretion discs \citep{Chandran2018} and may be the appropriate description of the heating mechanisms operating in our NF simulations. 

Kinetic processes beyond MHD may modify the fraction of energy radiated in the corona. For example, fast, collisionless magnetic reconnection or collisionless turbulent dissipation may act to more rapidly dissipate the Poynting flux launched into the corona, much of which we infer leaves in a magnetocentrifugal wind (Figure~\ref{fig:cooling_efficiency}). In addition, our models assume that all energy dissipated by the grid goes to heating the MHD fluid, i.e., the ions. In our simulations, this thermal energy can either be removed by cooling, or the energy can be advected out with the wind, particularly when $t_{\rm wind} < t_{\rm cool}$ (Figure~\ref{fig:timescales}). However, in reality, some of the dissipated energy should go directly to the leptons, not the ions; this energy will contribute to the radiation emitted by the corona. 
 
We close our discussion by noting that in defining the corona, we have chosen a fiducial transition height of $|z| = 2 H_z$. However, the height of this transition depends in reality on the lepton density distribution in the corona and the location of the $\tau_{\rm es} = 1$ surface. 

To see the connection between the lepton distribution and our two-temperature model, we can invert Equation~\ref{eq:A_Omega_teq} to determine a lepton number density $n_{l}$ for a given coronal ion number density $n_{i, \rm c}$, dimensionless electron temperature $\Theta_e$, and SMBH mass $M_{\rm BH}$,
\begin{equation}
    n_l \simeq 9 \times 10^{11} \: \mathcal{A} \left( \frac{n_{i, \rm c}}{n_0} \right) 
    \left( \frac{M_{\rm BH}}{10^7 \: \rm{M}_{\odot}} \right)^{-1} 
    \left( \frac{\Theta_e}{0.2} \right)^{3/2}
    \left( \frac{\ln{ \Lambda_e }}{23} \right)^{-1}.
\end{equation}
Because $n_l \propto \mathcal{A}$, our strongly cooled models ($\mathcal{A} \gtrsim 10^2$) have much larger lepton densities than their weakly cooled ($\mathcal{A} \lesssim 10^2$) counterparts for fixed $n_{i, \rm c}/n_0$. Thus, our $\mathcal{A} = 10$ models, which show the least amount of cooling in the corona, are likely radiating their energy at the lowest optical depths. Further, because $n_l$ increases with $n_{i, \rm c}/n_0$, strong winds such as those in the moderate NF simulations act to greatly increase the lepton density in the corona. Thus, the optically thick-to-thin transition may occur at much larger $|z|$ in these simulations, and the high efficiencies measured by assuming that the corona begins above $|z| = 2 H_z$ may be an over-estimate of the true fraction of energy radiated by optically thin plasma. Ultimately, radiation MHD simulations which can self-consistently calculate the $\tau_{\rm es} = 1$ surface are necessary to evaluate if NF fields are truly efficient at dissipating energy in optically thin plasma, or if they primarily heat optically thick material high above the disc. 

\section{Summary and conclusions} 
\label{sec:conclusion}

The hard X-ray coronae of luminous AGN and X-ray binaries are shaped by thermodynamic processes: the interplay of heating and cooling. Because the ion--electron equilibration timescale is much longer than the orbital timescales or Compton cooling timescales in AGN coronae, the plasmas powering coronae are two-temperature, with the ion temperature greatly exceeding that of electrons (\S\ref{sec:thermo_timescales}). 

In this paper, we have isolated and analyzed the role of two-temperature thermodynamics by introducing a simple, parameterized cooling function (\S\ref{sec:two_temp_model}; Equation~\ref{eq:cooling_function}) that models the effects of Coulomb-collision-mediated energy exchange from ions to rapidly Compton-cooled leptons (electrons and positrons). We showed that the Coulomb coupling parameter $\mathcal{A}$ in our model corresponds to the inverse of the dimensionless equilibration timescale, $(\Omega t_{\rm eq})^{-1}$ (Equation~\ref{eq:temp_equilibration}), which can be constrained by observations (Equation~\ref{eq:Omega_teq_obs}). 

Using stratified shearing-box simulations evolved with the \textit{Athena++} code (\S\ref{sec:methods}), we studied the effect of our two-temperature model on the vertical structure of a local patch of an accretion disc. The simplicity of our model allowed for a scan of three values of $\mathcal{A} \in \{ 10, 100, 10^3 \}$, and four different magnetic-field configurations, described by both the net vertical magnetic flux initially threading the disc, i.e. ``zero net flux'' (ZNF) vs. ``net flux'' (NF), and the initial midplane plasma beta parameter, $\beta_0$. In total, we ran~16 simulations that scanned field configuration and $\mathcal{A}$, including isothermal reference runs for each field configuration. We then analyzed the temperature, density, and magnetic-field structure (\S\ref{sec:flow_structure}; Figures~\ref{fig:volume_renderings}--\ref{fig:magnetization_alpha}) of our model coronae, and studied how outflows, both thermally driven (\S\ref{sec:thermally_driven_winds}) and magnetocentrifugally driven (\S\ref{sec:wind_accretion}), are affected by thermodynamics. 

Our main results are as follows:
\begin{enumerate}
    \item Our two-temperature models all form temperature inversions (Figure~\ref{fig:temperature_density}, left panels), with a hotter corona surrounding a colder disc. The coronal temperatures are a significant fraction of the virial temperature, and at least in the NF $\mathcal{A} = 10$, $\beta_0 = 10^4$ simulation, the peak horizontally averaged temperature remains a fixed fraction of the virial temperature, independent of box height (Appendix~\ref{sec:height_scan}). Using the virial temperature estimate from \S\ref{sec:thermo_timescales}, the NF $\mathcal{A} = 10$, $\beta_0 = 10^4$ simulation attains a coronal ion temperature of $6 \times 10^{10}$ K.
    \item Weaker cooling (smaller $\mathcal{A}$) tends to enhance the density of our coronae, and densities can rise by as much as an order of magnitude in ZNF simulations with $\mathcal{A} = 10$ compared to the isothermal reference simulation for the same field configuration at the same height. Still, the initial magnetic-field configuration has a greater effect on density structure than does thermodynamics, with NF simulations displaying higher densities due to strong magnetocentrifugally driven outflows (Figure~\ref{fig:temperature_density}, right panels).
    \item ZNF simulations exhibit weak, thermally driven outflows (Figure~\ref{fig:vz_profiles}, inset of left panel), although the slow magnetosonic point is only robustly within the domain for the $\mathcal{A} = 10$ runs. Outflows in NF simulations are all magnetocentrifugally driven, with the slow point well inside of the vertical boundary (Figure~\ref{fig:vz_profiles}, right panel).
    \item We find evidence for multiphase structure forming in our coronae, with broadened density and temperature distributions spanning more than 0.5~dex in log-density space and 5$T_0$ in temperature space respectively (\S\ref{sec:multiphase_corona}; Figures~\ref{fig:thermal_instability_ZNF}--\ref{fig:thermal_instability_NF}). Multiphase structure is strongly enhanced when the orbital timescale $t_{\rm orb}$, cooling timescale $t_{\rm cool}$, and the wind-outflow timescale $t_{\rm wind}$ obey the ordering, $t_{\rm orb} < t_{\rm cool} < t_{\rm wind}$ (Figure~\ref{fig:timescales}). This multiphase structure may alleviate problems with the `sandwich' corona by decreasing the covering fraction of the corona relative to the disc.
    \item While field-aligned thermal conduction is not included in our models, all of our coronae would be within the free-streaming regime of the field-aligned heat flux. Post-facto estimates of field-aligned free-streaming conduction indicate that conduction acts rapidly in these systems, on a timescale of a few to ${\sim} 10$ orbits (Figure~\ref{fig:fs_profiles}). The conduction time is, however, longer than the ion--electron equilibration time, $t_{\rm eq}$. Naively, we expect from this estimate that ion cooling due to Coulomb collisions is more important than conduction. 
    \item Wind mass- and energy-outflow rates decrease with increasing $\mathcal{A}$, although the mass-outflow rate depends more strongly on initial $\beta_0$ for NF fields than on $\mathcal{A}$ (Figure~\ref{fig:wind_mass_energy}). At least in the NF $\mathcal{A} = 10$, $\beta_0 = 10^4$ simulation, the wind mass-outflow rate is largely independent of box height (Figure~\ref{fig:height_scan}), unlike the isothermal case. This raises the fascinating possibility that disc winds associated with two-temperature coronae may evaporate the disc faster than it can accrete, leading to a transition from a thin disc to a RIAF.   Indeed, the wind depletion times we find are only ${\sim}100$ orbits (Table~\ref{table:simulations}), potentially much shorter than the viscous time for a standard thin disc.   It remains to be seen, however, if global simulations with more realistic physics find similar results.
    \item NF fields launch substantially stronger time-averaged Poynting fluxes into the corona, up to 10~per cent of the injected energy for $\beta_0 = 10^4$, and as much as 26~per cent for $\beta_0 = 10^3$, with a weak dependence on thermodynamics. ZNF fields transport no more than 5~per cent of the injected energy, independent of field strength and $\mathcal{A}$ (Figure~\ref{fig:cooling_efficiency}, left top panel).
    \item Accretion flows with NF, as represented by our NF simulations, can radiate a large fraction of their total luminosity at large heights above the midplane, with moderate NF field ($\beta_0 = 10^3$) simulations achieving coronal luminosities $\dot{E}_{\rm cool}^{\rm cor} \gtrsim 30$~per cent of the total flow luminosity, $\dot{E}_{\rm cool}$. The coronal density is larger in this case, however, so it is not clear that a larger fraction of the total power is radiated at lower optical depth $\tau_{\rm es}$.  
\end{enumerate}

The high cooling efficiencies enabled by NF fields are consistent with observational expectations, which indicate that AGN release a few tens of per cent of their bolometric luminosity via their optically thin coronae \citep{Svensson1994, Wang2004, Uzdensky2013}. These findings provide compelling motivation for further investigation into models of coronal heating that invoke mechanisms appropriate to NF fields, rather than focusing solely on the traditional models of coronal loops (\citealt{Galeev1979}; \citealt{Scepi2023} reached similar conclusions based on global simulations of thin, magnetically arrested discs). Determining self-consistently if energy is dissipated in the optically thin plasma characteristic of the corona requires full radiation transport calculations, which may modify the density and temperature structure found in our local accretion flows. 
%particularly in the thin disc. 

\section*{Acknowledgements}

CJB is grateful for the technical advice and support provided by Christopher White, Matthew Coleman, Xuening Bai, and Andrea Antoni, particularly in the early stages of this project. We thank Yan-Fei Jiang for stimulating conversations and for motivating our study. We are thankful to the referee for careful reading and suggestions, which improved the manuscript. EQ thanks Nico Scepi for valuable conversations. CJB especially is thankful for helpful discussions with Chris Reynolds, Andy Fabian, and Erin Kara regarding observations of AGN coronae and for sparking excitement towards studying this subject. CJB is supported by the National Science Foundation (NSF) Graduate Research Fellowship.  This work was supported in part by a Simons Investigator award to EQ from the Simons Foundation.
Computational resources for our simulations were provided by the Princeton Institute for Computational Science and Engineering (PICSciE) and the Office of Information Technology’s High Performance Computing Center at Princeton University.

\section*{Software}

The stratified shearing-box MHD simulations presented in this work were performed using the \textit{Athena++} code \citep{Stone2020}. Three-dimensional renderings were produced using the VisIt software package \citep{Childs_VisIt_An_End-User_2012}, which is supported by the Department of Energy with funding from the Advanced Simulation and Computing Program, the Scientific Discovery through Advanced Computing Program, and the Exascale Computing Project. Analysis was performed using \textit{numpy} \citep{numpy}, figures were produced through \textit{matplotlib} \citep{matplotlib}, and we used color maps/ schemes from the \textit{cmasher} package \citep{cmasher}.

\section*{Data availability}

Simulation data is available upon request to the corresponding author.

%\bibliographystyle{mnras} 
%\bibliography{corona.bib}

\begin{appendix} 

\section{Wind-driven accretion} \label{sec:wind_survival}

\begin{figure*}
\hbox{
\includegraphics[width=1.0\textwidth]{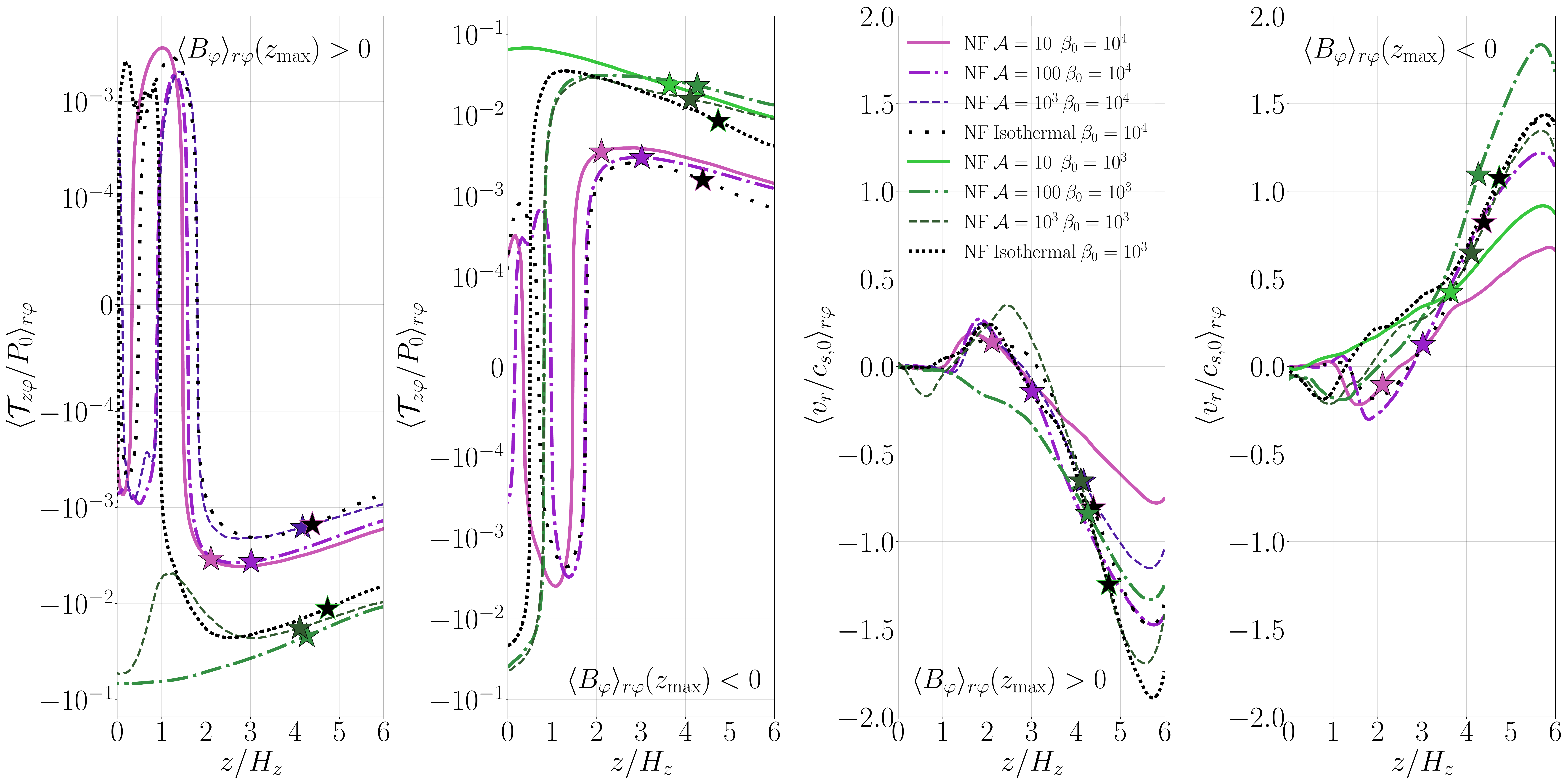}
}
\caption{Left: Horizontally averaged wind stress $\mathcal{T}_{z \varphi}$ averaged over dynamo cycles of the same sign in the toroidal magnetic field $B_{\varphi}$. Right: Horizontally averaged radial velocity $v_r$ averaged over dynamo cycles of the same sign in $B_{\varphi}$. The \alf point is represented by stars on each plot. Panels are shown separately for averages taken when the horizontally averaged $B_\varphi$ at the top boundary is positive ($\langle B_{\varphi} \rangle_{r \varphi} > 0$) or negative ($\langle B_{\varphi} \rangle_{r \varphi} < 0$), so as to avoid averaging out both $\mathcal{T}_{z \varphi}$ and $\rho v_r$, which change sign with the dynamo cycle.
}
\label{fig:Tzphi_profiles}
\end{figure*}

In \S\ref{sec:critical_points}, we discussed a criterion for assessing whether wind-driven accretion dominates over that driven by the radial turbulent transport of angular momentum, the latter quantified by $\alpha_{\rm mid}$. Here, we show that the wind stress $\alpha_z$ measured in our simulations depends on height because $\alpha_z$ acts to drive near-sonic radial flows at large $|z|$. By measuring $\alpha_z$ at the \alf point $z_{\rm A}$, we compute the critical disc thickness $(H_z/R_0)_{\rm crit}$ below which winds dominate the angular-momentum transport. By comparing the wind-driven accretion time, the viscous time, and the depletion time, we estimate a `survival thickness' for the disc, $(H_z/R_0)_{\rm surv}$, i.e., the thickness below which the accretion time is longer than the wind-depletion time such that the disc will evaporate to form a RIAF.

For this analysis, we assume that our local simulations represent a small patch of a much larger, global accretion disc. We assume the patch is centred at radius $R_0$, with an inner radius $R_{\rm in}$ and outer radius $R_{\rm out} = R_{\rm in} + \Delta R$, where $\Delta R/R_0 \ll 1$, such that the shearing-box approximation holds. Because $\langle \dot{M}_{\rm wind} \rangle_t$ and $z_{\rm A}$ depend only weakly on box height, at least for the NF $\mathcal{A} = 10$, $\beta_0 = 10^4$ simulation, we assume that the depletion times $t_{\rm dep}$ in our fiducial $z_{\rm max} = 6 H_z$ runs are representative of the time to deplete (i.e., remove all of the mass from) this small patch.

The equation of angular-momentum conservation in a global ideal MHD disc \citep{Balbus1998} is given by
\begin{equation} \label{eq:angular_momentum}
\begin{split}
     \frac{\partial}{\partial t} \left( \rho r v_{\varphi} \right) + \grad \bcdot r\left[ \rho u_{\varphi} \boldsymbol{u} - B_{\varphi} \boldsymbol{B}_p + \left(P + \frac{|\boldsymbol{B}_p|^2}{2}\right) \hat{\varphi} \right] = 0.
\end{split}
\end{equation}
Here, we work in global coordinates $(r, \varphi, z)$ such that $r$ represents the cylindrical radius, and the time evolution of the turbulent angular-momentum density $\rho r v_{\varphi}$ depends on the total velocity $\boldsymbol{u}$ as well as the poloidal magnetic field $\boldsymbol{B}_p$. The time-steady, $\varphi$-averaged angular momentum equation (\ref{eq:angular_momentum}) in a global disc is then
\begin{equation} \label{eq:phi_momentum}
    \frac{1}{r} \frac{\partial}{\partial r} r^2 \left(
    \langle \rho v_r \rangle_{\varphi} r \Omega + \langle \mathcal{T}_{r \varphi} \rangle_{\varphi} \right) + r \frac{\partial}{\partial z} \left( \langle \rho v_z \rangle_{\varphi} r \Omega + \langle \mathcal{T}_{z \varphi} \rangle_{\varphi} \right) = 0,
\end{equation}
where the wind-stress, i.e. the stress transporting $\varphi$-momentum (angular momentum) in the $z$-direction is 
\begin{equation}
    \mathcal{T}_{z \varphi} = \underbrace{\rho v_z v_{\varphi}}_{\rm Reynolds} - \underbrace{B_z B_{\varphi}}_{\rm Maxwell}.
\end{equation}
Using the $\varphi$-averaged equation of continuity (Equation~\ref{eq:mass}), we reduce Equation~\ref{eq:phi_momentum} to find
\begin{equation} \label{eq:radial_flows}
    \underbrace{\frac{1}{2} \langle \rho v_r \rangle_{r \varphi} \Omega}_{\rm radial \: flows} = \underbrace{\frac{1}{r^2} \frac{\partial}{\partial r} r^2 \mathcal{T}_{r \varphi}}_{\rm turbulent \: stress \: \sim \alpha} + \underbrace{\frac{\partial}{\partial z} \mathcal{T}_{z \varphi}}_{\rm wind \: stress \: \sim \alpha_z}.
\end{equation}
Physically, Equation~\eqref{eq:radial_flows} implies that radial flows (i.e. accretion flows) are driven both by the turbulent $r$--$\varphi$ stress provided by $\alpha$ and the wind stress provided by $\alpha_z \equiv \langle \mathcal{T}_{z \varphi}/P_0 \rangle_{r \varphi}$. 

A comparison of the turbulent stress and wind stress terms in Equation~\eqref{eq:radial_flows} provides a critical disc thickness $H_z/R_0$ below which, wind-driven accretion dominates, and above which, the turbulent ``viscosity'' afforded by $\alpha$ dominates,
\begin{equation} \label{eq:HoR_crit}
    \left( \frac{H_z}{R_0} \right)_{\rm crit} = \frac{\langle \mathcal{T}_{z \varphi} \rangle_{r \varphi}}{2 \langle \mathcal{T}_{r \varphi} \rangle_{r \varphi}}.
\end{equation}
The left panels of Figure~\ref{fig:Tzphi_profiles} show the horizontally averaged wind stress $\langle \mathcal{T}_{z \varphi} \rangle_{r \varphi}$ as a function of height $z \geq 0$ and the right panels of Figure~\ref{fig:Tzphi_profiles} show the corresponding horizontally averaged radial velocity $v_r$. Note that the sign of $\langle \mathcal{T}_{z \varphi} \rangle_{r \varphi}$ switches with dynamo cycles. Thus, similar to \cite{Fromang2013} who averaged $\mathcal{T}_{z \varphi}$ over a single dynamo cycle, we show the average of $\mathcal{T}_{z \varphi}$ over all dynamo cycles where $\langle B_{\varphi} \rangle_{r \varphi}$ is positive at the upper boundary separately from $\mathcal{T}_{z \varphi}$ averaged over all dynamo cycles where $\langle B_{\varphi} \rangle_{r \varphi}$ is negative at the upper boundary. We only show $\langle \mathcal{T}_{z \varphi} \rangle_{r \varphi}$ for $z \geq 0$.

Clearly, $\langle \mathcal{T}_{z \varphi} \rangle_{r \varphi}$ changes with $z$ such that we must make a choice of where to measure $\langle \mathcal{T}_{z \varphi} \rangle_{r \varphi}$ in order to compute the critical thickness with Equation~\ref{eq:HoR_crit}. This choice of measurement height is then necessarily arbitrary, precluding a clean measurement of $(H_z/R_0)_{\rm crit}$. The vertical gradient in the wind stress also drives significant radial flows in the domain, particularly in the surface layers of the disc. Examining the horizontally averaged $v_r$ profiles in Figure~\ref{fig:Tzphi_profiles}, we see that in all but the NF $\mathcal{A} = 10$ $\beta_0 = 10^4$ simulation, radial flows are supersonic (with respect to the midplane sound speed, $c_{s,0}$) above $z = (4-5) H_z$. 

Since the wind's lever arm is determined by the height of the \alf point, we choose to measure the wind stress at the \alf point $z_{\rm A}$, as indicated by stars in Figure~\ref{fig:Tzphi_profiles}. Measuring $\langle \mathcal{T}_{r \varphi} \rangle_{r \varphi}$ at the midplane and $\langle \mathcal{T}_{z \varphi} \rangle_{r \varphi}$ at $z_{\rm A}$, we find that the critical thickness is $(H_z/R_0)_{\rm crit} = 0.02 - 0.04$ for all NF simulations. Thus, angular momentum transport is dominated by winds only for discs thinner than $H_z/R_0 \approx 0.02$.

To assess the role of wind-driven depletion on the survival of the thin-disc, we compare the viscous time ($t_{\rm visc}$) and the depletion timescale ($t_{\rm dep}$; Equation~\ref{eq:tdep}). The viscous time is related to the thermal time via $t_{\rm visc} = (R_0/H_z)^2 t_{\rm thm}$ such that the viscous time is equal to the depletion time when
\begin{equation} \label{eq:HoR_surv}
    \left( \frac{H_z}{R_0} \right)_{\rm surv} \approx \left( \frac{t_{\rm thm}}{t_{\rm dep}} \right)^{1/2}.
\end{equation}
This relation implies that $(H_z/R_0)_{\rm surv} \approx 0.1$ for our NF simulations (based on the measurements of $t_{\rm thm}$ and $t_{\rm dep}$ from Table~\ref{table:simulations}).  Since at this $H_z/R_0$ the internal stresses indeed dominate over wind stresses in our simulations, it is self-consistent to have focused on the internal viscous stresses in the above estimate.

Discs thinner than $(H_z/R_0)_{\rm surv} \approx 0.1$ will be depleted by the wind, and the depletion process may prompt a runaway evaporation of the thin disc. Note that our calculation of $(H_z/R_0)_{\rm surv}$ assumes that the depletion time in our local simulations is the same as would be the case in a global accretion disc. This assumption is motivated by the fact that mass-outflow rates, and thus depletion times, in our two-temperature models show better convergence with box height compared to isothermal calculations (at least for our NF $\mathcal{A} = 10$ $\beta_0 = 10^4$ simulations). Ultimately, global calculations are required to address the issues of wind depletion and disc evaporation. We defer the resolution of these questions to future work.

\section{Box-height Scan} \label{sec:height_scan}

\begin{figure*}
\hbox{
\includegraphics[width=1.0\textwidth]{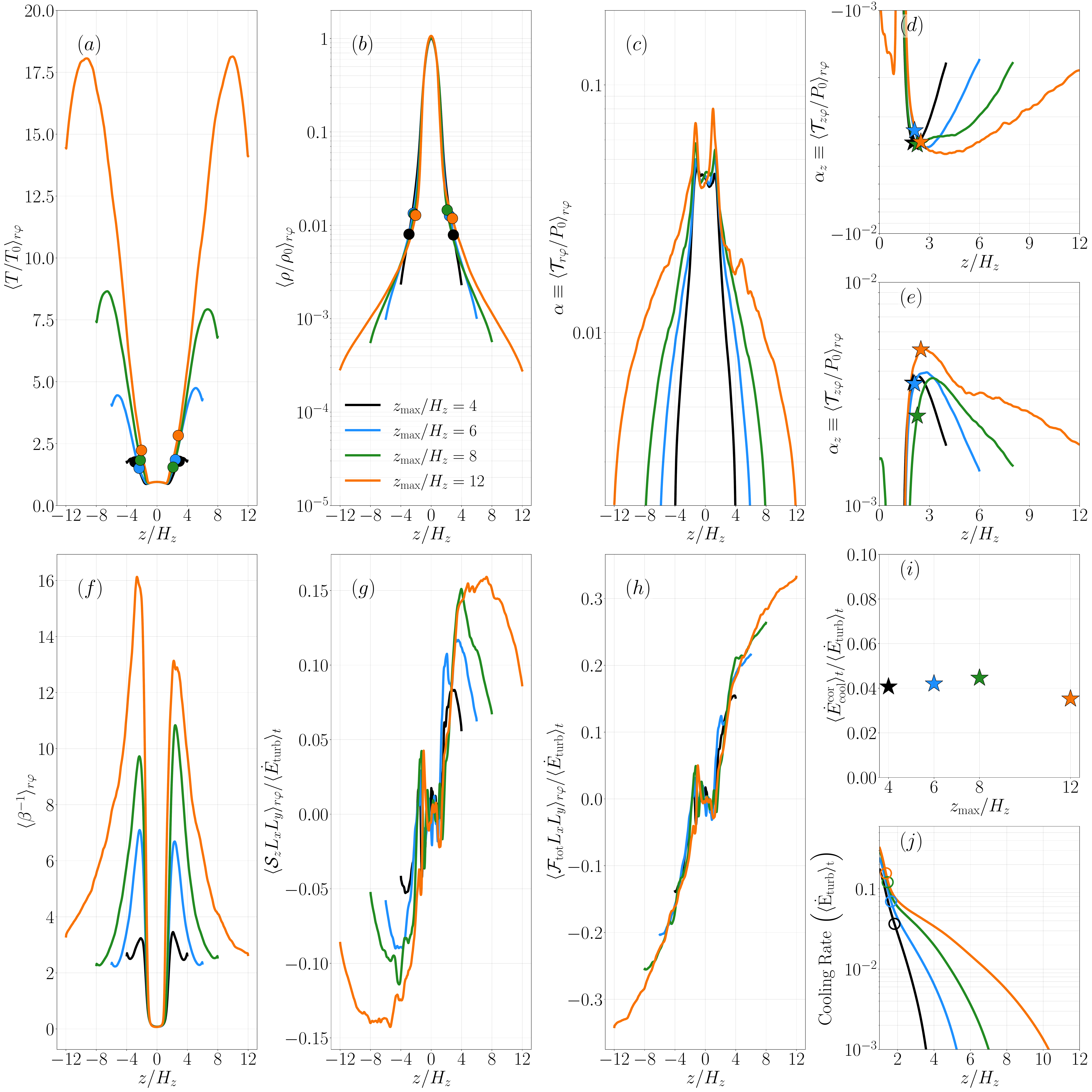}
}
\caption{Horizontally averaged profiles for the scan of box height $z_{\rm max}$ for the net-flux (NF) $\mathcal{A} = 10$ $\beta_0 = 10^4$ simulations. We show profiles of (a) temperature, (b) density, (c) turbulent stress $\alpha$ parameter, (d) wind stress $\alpha_z$ for $\langle B_{\varphi} \rangle_{r \varphi} > 0$ at the top boundary, (e) wind stress $\alpha_z$ for $\langle B_{\varphi} \rangle_{r \varphi} < 0$ at the top boundary, (f) magnetization parameter ($\beta^{-1}$; Equation~\ref{eq:beta_inv}), (g) Poynting flux ($\mathcal{S}_z$; Equation~\ref{eq:Poynting_flux}), and (h) total vertical energy flux $\mathcal{F}_z$. All profiles are averaged from 30--100 orbits, except in panels (g), (h), (i), and (j), which are averaged from 30--60 orbits. The different time-frame is chosen because of an asymmetry which arises in the toroidal field dynamo at 60 orbits in the tallest box run. In panel (i), we show the total cooling power above $|z| = 2 H)z$ as a function of box height, and panel (j) shows the cooling rate as a function of height $z$ above the disc, similar to the panels on the right of Figure~\ref{fig:cooling_efficiency}.
}
\label{fig:height_profiles}
\end{figure*}

Figure~\ref{fig:height_profiles} shows the results obtained throughout this paper for our box-height scan of the NF $\mathcal{A} = 10$ $\beta_0 = 10^4$ simulations. The panels in Figure~\ref{fig:height_profiles} are lettered, and we state the main takeaway points in the order corresponding to the lettered panels.
\renewcommand{\labelenumi}{\alph{enumi})}
\begin{enumerate}
    \item The peak temperature in the corona scales as $z_{\rm max}^2$, which is the same scaling as the virial temperature, $T_{\rm virial}$ (Equation~\ref{eq:T_virial}). Increasing $z_{\rm max}$ causes $z_T$ to move inward slightly, although temperature profiles are relatively converged with box height, at least until the profiles turn over at higher $|z|$, which is an effect of the boundary.
    \item Density profiles are converged with box height. 
    \item Profiles of $\alpha$ show the same value at the midplane for different box heights, although there is a clear enhancement of the accretion rate in the disc surface layers. We take this enhancement, concentrated in a narrow range around $|z| \approx H_z$, to be evidence of magnetically elevated accretion, and taller boxes show greater enhancement. There are clear wings in the $\alpha$ profiles, which increase dissipation in the corona and the total injected energy, $\langle \dot{E}_{\rm turb} \rangle_t$.
    \item The wind stress $\alpha_z$ is converged with box height at both the \alf point and the boundary. 
    \item For dynamo cycles with $\langle B_{\varphi} \rangle_{r \varphi} < 0$ at the upper boundary, $\alpha_z$ profiles show modest variations with box height. However, the primary uncertainty in calculating the wind stress relative to the turbulent stress via  Equation~\ref{eq:HoR_crit} is the question of where to measure $\alpha_z$ for comparison with $\alpha_{\rm mid}$.
    \item The magnetization parameter $\langle \beta^{-1} \rangle_{r \varphi}$ profiles are converged with height until the point where the profiles turn over higher in the corona. This turnover, like the turnover in temperature, is likely a boundary effect. The peak magnetization of the coronae in taller boxes is larger such that the coronae formed in taller boxes are more strongly magnetized.
    \item Vertical Poynting flux $\mathcal{S}_z$ profiles are converged with box height until the turnover point caused by the boundary. Poynting flux continues to increase with height until the turnover point, indicating that assessing the Poynting flux into the `corona' depends on the height of the coronal transition. In a radiation-MHD simulation, this height is the $\tau_{\rm es} = 1$ surface.
    \item The total energy flux profiles are converged with height but continue to increase with $|z|$ out to the boundary. Because the fast point is never on the domain for any simulation, final mass energy outflow rates depend on box height.
    \item The total cooling rate in the corona is converged with box height.
    \item Cooling rate profiles (i.e. cooling rates as a function of $|z|$) in the corona are \textit{not} converged with box height. Variations in the cooling rate profiles are likely a result of the wings in the $\alpha$ parameter profiles (panel (c)). Cooling rates are relatively similar out to $|z| = 2 H_z$, but higher in the corona, taller boxes show a larger cooling rate at fixed $|z|$. This excess cooling has a negligible effect on the total cooling power emitted by the corona, which is dominated by denser regions closer to the disc.
\end{enumerate}

\end{appendix}

% required because of bug in MN2e style file
% throws away figs otherwise
\clearpage

\end{document}